\documentstyle[prc,aps,preprint,psfig]{revtex}
\draft
\begin{document}

\title{Open charm production in relativistic nucleus-nucleus collisions
\thanks{supported by  GSI Darmstadt and FZ J\"ulich}  }

\author{W. Cassing, E. L. Bratkovskaya, A. Sibirtsev \\[5mm]
{\normalsize Institut f\"{u}r Theoretische Physik,}\\ {\normalsize
Universit\"{a}t Giessen,} {\normalsize 35392 Giessen, Germany} }
\maketitle

\begin{abstract}
We calculate excitation functions for open charm mesons in $Au+Au$
reactions from AGS to RHIC energies within the HSD transport
approach which is based on string, quark, diquark ($q, \bar{q},
qq, \bar{q}\bar{q}$) and hadronic degrees of freedom. The open
charm cross sections from $pN$ and $\pi N$ reactions are fitted to
results from PYTHIA and scaled in magnitude to the available
experimental data. From our dynamical calculations we find an
approximate $m_T$-scaling for pions, kaons, $D$-mesons and
$J/\Psi$ -- when discarding final state elastic scattering of
kaons and $\phi$-mesons with pions -- in central collisions of $Au
+ Au$ at 160 A$\cdot$GeV (with an apparent slope of 176 MeV)
without employing the assumption of a Quark-Gluon Plasma (QGP). We
demonstrate that this result is essentially due to a relative
$m_T$-scaling in $pp$ collisions at $\sqrt{s} \approx$ 17.3 GeV.
At lower bombarding energies of 25 A$\cdot$GeV a suppression of
$D$-mesons by a factor of $\sim$ 10 relative to a global
$m_T$-scaling with slope 143 MeV is expected. However, when
incorporating attractive $D$-meson self energies as suggested by
QCD sum rules, an approximate $m_T$-scaling is regained even at 25
A$\cdot$GeV. The effects of $D$-meson rescattering and charmonium
absorption are discussed, furthermore, with respect to rapidity
and transverse mass distributions in central collisions of $Au +
Au$ at 25, 160 A$\cdot$GeV and 21.3 A$\cdot$TeV.
\end{abstract}

\noindent
\vspace{10mm} \noindent


\narrowtext
\newpage

\section{Introduction}

Apart from the light and strange flavor
($u,\bar{u},d,\bar{d},s,\bar{s}$) quark physics and their hadronic
bound states in the vacuum ($\pi, K, \phi$ etc.) the interest in
hadronic states with charm flavors ($c, \bar{c}$) has been rising
continuously in line with the development of new experimental
facilities \cite{QM96,QM97,QM99,SQM99}. This relates to the charm
production cross section in $pN$ and $\pi N$ reactions as well as
to their interactions with baryons and mesons which determine
their properties (spectral functions) in the hadronic medium.

The charm quark degrees of freedom have gained vivid interest
especially in the context of a phase transition to the quark-gluon
plasma (QGP) \cite{Heinz} where $c\bar{c}$ meson states should no
longer be formed due to color screening \cite{Satz,Satznew}.
However, the suppression of $J/\Psi$ and $\Psi^\prime$ mesons in
the high density phase of nucleus-nucleus collisions
\cite{NA50b,NA50a} might also be attributed to inelastic comover
scattering (cf.
\cite{Cass99,Vogt99,Gersch,Seattle98,Cass00,Capella} and Refs.
therein) provided that the corresponding $J/\Psi$-hadron cross
sections are in the order of a few mb
\cite{Haglin,Haglin2,Konew,Ko,Sascha,Sascha2}. Present theoretical
estimates here differ by more than an order of magnitude
\cite{Bernd} especially with respect to $J/\Psi$-meson scattering
such that the question of charmonium suppression is not yet
settled. On the other hand, the enhancement of 'intermediate-mass
dileptons' in $Pb + Pb$ collisions at the SPS has been tentatively
attributed to an enhancement of 'open charm' in nucleus-nucleus
collisions relative to $pA$ reactions at the same invariant energy
$\sqrt{s}$ \cite{NA50d}. It should be mentioned that this
enhancement does not stem from the charmonium dissociation since
it is about two orders of magnitude larger the total charmonium yield.
Thus 'charmonium suppression' and 'open
charm enhancement' are present facets of relativistic heavy-ion
collisions.

Furthermore, it is well known experimentally \cite{PDG} that the
$D,\bar{D}$ and $D^*, \bar{D}^*$ mesons show some analogy to the
$K, \bar{K}$ and $K^*, \bar{K}^*$ mesons with respect to their
excitation spectrum because the strange (antistrange) quark is
replaced by a charm (anticharm) quark in the hadronic state. Since
quite substantial in-medium potentials have been suggested for
antikaons in dense nuclear matter \cite{Cass99,Brown}, the latter
might also show up for the corresponding $D$-mesons in view of a
similar wavefunction for the light quark \cite{Alex99}. In fact,
QCD sum rule studies point towards attractive potentials for the
$D$-mesons \cite{Weise} which might lead to enhanced production
cross sections of open charm especially at low bombarding energies
close to threshold. Substantially lower in-medium effects are expected
for the $J/\Psi$ or $\eta_c$ do to a small coupling of the $c, \bar{c}$
quarks to the nuclear medium \cite{Weise2}. Thus the charm-meson sector, which is
insufficiently known so far, provides a theoretical
\cite{Ko,Johanna,Peter,Rafelski,Wang,Lin00} and experimental
challenge for the future \cite{NA50new,NA49new}.

In this work we will explore the perspectives for open charm
production in nucleus-collisions from AGS to RHIC energies
employing the HSD transport approach \cite{Cass99,Cass00} for the
overall reaction dynamics using parametrizations for the
elementary production channels including the charmed hadrons $D,
\bar{D}, D^*, \bar{D}^*, D_s, \bar{D}_s, D_s^*, \bar{D}_s^*,$
$J/\Psi, \Psi(2S), \chi_{2c}$ from $NN$ and $\pi N$ collisions.
The latter parametrizations are fitted to PYTHIA calculations
\cite{PYTHIA} above $\sqrt{s}$ = 10 GeV and extrapolated to the
individual thresholds, while the absolute strength of the cross
sections is fixed by the experimental data
\cite{NA16,NA27,E743,E653,E789,NA32,E653b,NA32b,E769,WA92,E791}
similar to Ref. \cite{Peter} (Section 2). The production of open
charm in central collisions of $Au + Au$ at 25, 160 A$\cdot$GeV
and 21.3 A$\cdot$TeV is studied in Section 3 with respect to
transverse mass ($m_T$) and rapidity distributions. We will first
switch off elastic collisions of kaons and $\phi$-mesons with
pions in nucleus-nucleus collisions to allow for a more
transparent comparison to the spectra from $pp$ collisions (Sect.
3.1 -- 3.3). The modifications of the $m_T$-spectra due to the
latter elastic interactions will be discussed in Section 3.4 while
an excitation function for various mesons in central $Au + Au$
collisions is presented in Section 3.5. Section 4 concludes this
study with a summary and discussion of open problems.

\section{Elementary cross sections from $pN$ and $\pi N$
collisions}

Before examining nucleus-nucleus collisions we have to specify the
differential open charm cross sections from $pN$ and $\pi N$
reactions that will enter the HSD approach. Contrary to light
meson production in hadronic reactions the creation of a
$c\bar{c}$ pair is due to a hard process and dominated by
gluon-gluon fusion at high $\sqrt{s}$. Using MRS G (next to
leading order) structure functions from the PDFLIB package
\cite{MRS} for the gluon distribution of the proton, a bare charm
quark mass $m_c=$ 1.5 GeV and $k_T$ = 1 GeV we obtain the cross
sections for $D, \bar{D}, D^*, \bar{D}^*, D_s, \bar{D}_s, D_s^*,
\bar{D}_s^*,$ as a function of $\sqrt{s}\geq $ 10 GeV from PYTHIA
\cite{PYTHIA} as displayed in Fig. 1 (upper part). Corresponding
results for $\pi N$ reactions are shown in the lower part. Since
the individual lines are hard to distinguish, some general trends
are pointed out:  All cross sections indicate a common (smooth)
energy dependence. The $D^*$-mesons are created more abundantly
than the $D$-mesons roughly by a factor of 3 due to the three
different spin polarizations; the small mass difference between
$D$- and $D^*$-mesons of $\approx$ 140 MeV plays almost no role
for $\sqrt{s} \ge$ 10 GeV. On the other hand, an exchange of a
light ($u,d$) quark by a strange ($s$) quark costs a factor of
3-4. Consequently, the cross sections of $D^0, \bar{D}^0, D^+,
D^-$ and $D_s^*, \bar{D}_s^*$ are roughly comparable at high
$\sqrt{s}$. These simple considerations specify the relative
abundance of the open charm mesons. However, the absolute
magnitude of the cross sections is not expected to match
experimental data due to the perturbative nature of these
calculations and rescaling factors $K$ have to be introduced
\cite{Peter}.

In this spirit we fit the individual results from PYTHIA
(multiplied by factors of 12 and 7 for $pN$ and $\pi N$,
respectively) by an expression of the form,
\begin{equation}
\sigma_X(s) = a_X (1 - Z)^\alpha \ Z^{-\beta},
 \label{fit}
\end{equation}
with $Z = {\sqrt{s^0_X}}/{\sqrt{s}}$ where $\sqrt{s^0_X}$ denotes
the threshold for the channel $X$ in $pN$ or $\pi N$ reactions.
Note that close to threshold the production for
$\bar{D}(\bar{c})$-mesons is enhanced relative to $D(c)$-mesons
since the $c$-quark can end up in $\Lambda_c, \Sigma_c,
\Sigma_c^*$ baryons with lower threshold, while $D(c)$-mesons
require the associated production with a $\bar{D}(\bar{c})$ meson.
These threshold phenomena are in close analogy to the strangeness
sector, where the mesons with a $\bar{s}$-quark are produced close
to threshold essentially together with hyperons ($\Lambda,
\Sigma$), whereas antikaons require the associated production with
a kaon.

The formula (\ref{fit}) ensures the proper thresholds by construction
while the exponents $\alpha$ and $\beta$ describe the rise at threshold
and the asymptotic behaviour, respectively. In order to properly
'normalize' the results from Fig. 1 we address to the experimental 
data from Refs.
\cite{NA16,NA27,E743,E653,E789,NA32,E653b,NA32b,E769,WA92,E791} that
have been extrapolated to full open charm cross sections by using the
charge ratio's as given by PYTHIA. Furthermore, we have used a factor
of 2 when extrapolating data for $x_F > 0$ to the full Feynman $x_F$
regime for $pN$ collisions and a factor of 1.6 for $\pi N$ reactions
\cite{Frixione}. The results of the fits are given in Tables 1 and 2
for the parameters $a_X, \alpha$ and $\beta$. We mention that the high
value of the exponent $\alpha$ compared to related fits for $\rho$,
$\omega$ or $\phi$ production \cite{Cass99} indicates the different
production mechanism for $c\bar{c}$ pairs compared to light quark
pairs.

The parametrized results from this extrapolation are displayed in
Fig. 2 for $pN$ (upper part) and $\pi N$ reactions (lower part)
for the full charm cross section including all mesons as specified
above with their individual thresholds. The solid lines in Fig. 2
represent the sum over all open charm mesons (within the
parameters given in Tables 1 and 2) while the individual lines refer to the
individual mesons that are somewhat hard to disentangle.
As in Fig. 1 these cross sections group to 3 bunches at
high $\sqrt{s}$ where the upper bundle of lines corresponds to
$D^{*+},D^{*-}, D^{*0}$ and $\bar{D}^{*0}$, the middle bundle to
$D^{+},D^{-}, D^{0}, \bar{D}^{0}$ and the vector states with a
strange quark $D_s^{*}, \bar{D}_s^*$, while the lower bundle gives
the cross section for $D_s, \bar{D}_s$.

It is interesting to compare these results with the cross sections
for $J/\Psi$ (including $\chi_c$ decay) and $\Psi^\prime \
(\Psi(2S))$ which are displayed in Fig.  3 as a function of
$\sqrt{s}$ together with the parametrizations (solid lines) for
$pN$ and $\pi N$ reactions (taken from Ref.
\cite{Vogt99,schuler}).  Note that at $\sqrt{s}$ = 20 GeV open
charm is enhanced by about a factor of 50 relative to $J/\Psi$ and
that this ratio increases  with the available energy. Since the
parametrization from Ref. \cite{schuler} approaches some constant
value at high $\sqrt{s}$ contrary to the PYTHIA calculations (cf.
Fig. 1 of Ref. \cite{Vogt99}) we have fitted the total cross
section by the function
\begin{equation}
\sigma_X(s) = b_X (1 - Y)^\alpha \ Y^{-\beta} \ \Theta(\sqrt{s}-\sqrt{s_0})
 \label{fitj}
\end{equation}
with $Y=m_X/\sqrt{s}$ and $\alpha$ = 10, while $\sqrt{s_0}$
denotes the threshold in vacuum. Again the parameter $\beta$
governs the high energy rise of the cross section which for $\beta
\approx$  1 is now in line with the PYTHIA calculations specified
above. Our fits give $b_{J/\Psi}$ = 96 nb, $b_{\chi_c}$ = 64 nb,
$b_{\Psi^\prime}$ = 20 nb; the results for $J/\Psi$ (including the
$\chi_c$ decay) are shown in the upper part of Fig. 3 in terms of
the dashed line.

The cross sections (\ref{fit}),(\ref{fitj}) will be used in the
transport calculations to be discussed below which, apart from the
total cross sections, also need the differential distribution of
the produced mesons in the transverse momentum $p_T$ and the
rapidity $y$ (or Feynman $x_F$) from each individual collision. We
recall that $x_F = p_z/p_z^{max} \approx 2 p_z/\sqrt{s}$ with
$p_z$ denoting the longitudinal momentum. For the differential
distribution in $x_F$ and $p_T$ we use the ansatz,
\begin{equation}
\frac{1}{2 p_T} \frac{dN}{dx_F dp_T} \sim (1 - |x_F|)^\gamma \ \exp(-b p_T),
\label{fit2}
\end{equation}
with $\gamma \approx $ 4.5 and $b \approx$ 3.0 GeV$^{-1}$. With
these parameters the differential transverse momentum
distributions of $D/\bar{D}$ mesons in $pp$ (and $\pi N$)
reactions at 250 GeV \cite{E769} may reasonably be described as
shown in Fig. 4. The $x_F$ and $p_T$ distribution for charmonium
production, furthermore, is taken from Ref. \cite{Rvogt}.

We have to point out that our parametrizations for the
differential and total cross sections for open charm (as well as
charmonia) become questionable at low energy, but also at high
energy. It is thus mandatory that they have to be controlled by
experimental data from $pp$, $pA$ and $\pi N$ reactions before
reliable conclusions on open charm dynamics in nucleus-nucleus
reactions can be drawn.

For the interpretation of the results from nucleus-nucleus
collisions (cf. Section 3) it is worth to compare to $pp$
collisions at different energies, respectively. To this aim we
display in Figs. 5--7 the differential multiplicities $(2
m_T)^{-1} dN_X/dm_T$ in the transverse mass
\begin{equation}
m_T = \sqrt{p_T^2 + m_X^2} \end{equation}
 for all final pions,
kaons, $\phi$-mesons, $D + \bar{D}$ mesons and charmonia from $pp$
reactions at $\sqrt{s}$ = 7.1 GeV, 17.3 GeV and 200 GeV,
respectively. The pion spectra describe the sum of $\pi^+, \pi^0,
\pi^-$, the kaon spectra the sum of $K^+, K^0, \bar{K}^0, K^-$,
the $D$-meson spectra the sum of all $D, D^*,D_s,D_s^*$ and their
antiparticles while the spectrum denoted by $c\bar{c}$ includes
the $J/\Psi$, the $\chi_c$ as well as the $\Psi^\prime$, where the
latter contribution starts at $m_T \approx 3.7$ GeV and becomes
visible as a tiny kink in the $m_T$-spectra. Here the open charm
and charmonia results stem from the parametrizations specified
above (including the decay $\chi_c \rightarrow J/\Psi + \gamma$)
while the spectra for pions, kaons and $\phi$-mesons are from the
LUND string model \cite{LUND} (as implemented in the HSD transport
approach). For orientation we also show exponential spectra with
slope parameters of 143 MeV, 176 MeV and 225 MeV, respectively,
which describe the $m_T$-spectra of pions rather well. The kaon
spectra at all energies are down by a factor of $\sim$ 3, the
$\phi$ spectra by a factor of 9-10 relative to this line due to
strangeness suppression in $pp$ collisions.  However, it is quite
remarkable that the charmonia spectra fit well to this approximate
$m_T$-scaling (within a factor of 2-3) at $\sqrt{s}$ = 7.1, 17.3
and 200 GeV, respectively. Furthermore, the spectrum of open charm
is roughly compatible with $m_T$-scaling at $\sqrt{s}$ = 17.3 and
200 GeV, while the $D, \bar{D}$ mesons are suppressed relative to
the scaling by a factor $\sim$ 30 close to threshold ($\sqrt{s}$ =
7.1 GeV). Whereas these results basically stem from our
parametrizations at $\sqrt{s}$ = 7.1 and 200 GeV, the spectra at
$\sqrt{s}$ = 17.3 are controlled by experimental data. Such an
'apparent' statistical production of mesons in elementary
reactions has been advocated before by Becattini \cite{Becattini}.

We point out that also the approximate $m_T$-scaling from Figs.
5--7 has to be controlled by explicit experimental measurements.
Data in a limited rapidity range might lead to somewhat different
results since the rapidity distributions of pions, kaons,
$\phi$'s, $D$'s, $D^*$'s and charmonia differ substantially due to
kinematical reasons, i.e. the width of the rapidity distribution
decreases with increasing meson mass.

\section{Nucleus-nucleus collisions}

Inspite of the inherent uncertainties pointed out above it is
worthwhile to explore the dynamics of open charm mesons in
relativistic nucleus-nucleus collisions. Experiments are planned
at the SPS \cite{NA50new} as well as at RHIC and might be even
performed in the 20--30 A$\cdot$GeV region \cite{GSI}. Here we
will employ the HSD transport approach for the nucleus-nucleus
dynamics that has been tested in detail for $pp$, $pA$ and $AA$
reactions from SIS to SPS energies \cite{Cass99,Ko95,Geiss} and
been used for the description of charmonium production and
propagation as well \cite{Cass00,Cass97,Geiss99}.

We recall that (as in Refs. \cite{Cass97,Geiss99}) the charm
degrees of freedom are treated perturbatively and that initial
hard processes (such as $c\bar{c}$ or Drell-Yan  production from
$NN$ collisions) are 'precalculated' to achieve a scaling of the
inclusive cross section with the number of projectile and target
nucleons as $A_P \times A_T$. To implement this scaling we
separate the production of the hard and soft processes: The
space-time production vertices of the $c\bar{c}$ pairs are
calculated in each transport run by neglecting the soft processes,
i.e. the production of light quarks and assosiated mesons. The
resulting number $N_{coll}(b)$  of these 'hard' collisions is
shown for $Au + Au$ at 160 A$\cdot$GeV in Fig. 8 (full squares) as
a function of impact parameter. The inclusive number of inelastic
$NN$ collisions is given by the integral of $N_{coll}$ over impact
parameter
\begin{equation}
I = \frac{2 \pi \int  b \ N_{coll}(b) \
db}{\sigma_{inel.}(\sqrt{s})} \approx A^2, \label{int1}
\end{equation}
which gives approximately $A^2$, i.e. the experimental scaling for
'hard' processes.  In (\ref{int1}) the mass number $A = 197$ for
$Au$, while $\sigma_{inel.}(\sqrt{s}) (\approx 34$ mb) denotes the
inelastic nucleon-nucleon cross section. The calculated
$N_{coll}(b)$ compares well with the result from Glauber theory
(solid line in Fig. 8), where the number of inelastic interactions
in nucleus-nucleus collision $A{+}B$ at impact parameter ${\bf b}
= (b,0,0)$ is given as~\cite{Formanek,Czyz}
\begin{equation}
N_{AB}(b) = A\, B\int \sigma_{inel}\, T_A({\bf s}) \, T_B({\bf
s}-{\bf b }) \, d^2s, \label{Glauber}
\end{equation}
where ${\bf s} = (s_x,s_y,0)$ is orthogonal to the
$z$-(beam-)direction. In the integral (\ref{Glauber})
\begin{equation}
T_A (b)=\intop_{-\infty}^{+\infty}\rho (\sqrt{b^2+z^2}) \, dz
\end{equation}
is the profile function normalized to unity, while  $\rho ({r})$
is the nuclear density taken of Woods-Saxon shape.

Thus the scaling for initial hard processes is adequately realized
in the transport approach.  We mention that this scaling
prescription might no longer be valid at low and high energy due
to modifications of the gluon structure functions during the
heavy-ion reaction or related shadowing phenomena \cite{Strikman}.
For our initial study, however, we discard such effects.

Apart from primary hard $NN$ collisions the open charm mesons or
charmonia may also be generated by secondary 'meson'-'baryon'
reactions. Here we include all secondary collisions of mesons with
'baryons' by assuming that the open charm cross section (from
Section 2) only depends on the invariant energy $\sqrt{s}$ and not
on the explicit meson or baryon state. Furthermore, we take into
account all interactions of 'formed' mesons -- after a formation
time of $\tau_F$ = 0.8 fm/c (in their rest frame) \cite{Geiss} --
with baryons or diquarks, respectively.

In the transport calculation we follow the motion of the
charmonium pairs or produced $D, \bar{D}$-mesons within the full
background of strings/hadrons by propagating them as free
particles, i.e. neglecting in-medium potentials\footnote{Except
for the case of in-medium mass shifts in Section 3.2}, but follow
their collisional history with baryons and mesons or quarks and
diquarks. For reactions with diquarks we use the corresponding
reaction cross section with baryons multiplied by a factor of 2/3.
For collisions with quarks (antiquarks) we adopt half of the cross
section for collisions with mesons. Whereas the latter concept is
oriented at the additive quark model, this assumption still does
not solve the problem since the cross sections of $D$-mesons or
charmonia with baryons and various mesons (essentially $\pi$,
$\rho$ and $\omega$ mesons) are not well known. Thus we will
provide results with and without rescattering of open charm
mesons.

In order to study the effect of rescattering we tentatively adopt
the following dissociation cross sections of charmonia with
baryons independent on the energy:
\begin{equation}
\sigma_{c\bar{c}B} = 6 \ {\rm mb}; \ \sigma_{J/\Psi B} = 4 \ {\rm mb}; \
\sigma_{\chi_c B} = 5 \ {\rm mb}; \  \sigma_{\Psi^\prime B} = 10 \ {\rm mb},
\label{sigmacB}
\end{equation}
 while a lifetime (in it's rest frame) of 0.4 fm/c is
assumed for the pre-resonance $c\bar{c}$ pair \cite{Kharz}. The
energy-dependent $J/\Psi$-meson cross sections for dissociation to
$D\bar{D}$ are taken from the calculations of Haglin \cite{Haglin}
which on average lead to a similar $J/\Psi$ comover suppression
than the overall cross section of 3 mb adopted in Ref.
\cite{Cass97}.

On the other hand, the $D/\bar{D}$ mesons are expected to have
large cross sections with mesons or baryons due to the light
flavor content such that light meson ($\pi, \rho, \omega, \eta$)
exchanges should describe the dominantly elastic cross sections at
low relative momenta. We here adopt the calculations from Ref.
\cite{Ko} which predict elastic cross sections in the range of
10--20 mb for $D, D^*$ scattering with mesons dependent on the
size of the formfactor employed. As a guideline we use a constant
cross section of 10 mb for elastic scattering with mesons and also
baryons, although the latter might be even higher for very low
relative momenta. We neglect charm exchange reactions such as $D^+
N \rightarrow \pi \Lambda_c, \Sigma_c$ or $\pi \Lambda_c
\rightarrow \bar{D} N$ in the present study, which will
essentially modify the charm quark content of mesons relative to
baryons. Furthermore, we discard a recreation of charmonia by
channels such as $D+ \bar{D} \rightarrow J/\Psi + \pi$, since at
AGS and SPS energies these reactions are negligible \cite{Peter};
on the other hand, at RHIC energies this charmonium formation
might become essential \cite{Rafelski}. However, the formation
cross sections are not well known and the significance of these
channels is discussed controversely in the present literature
\cite{Johanna,Rafelski,Redlich}.

In the transport calculations to be discussed below we will focus
on the relative yield of pions, kaons, $\phi$-, $D+\bar{D}$-mesons
and charmonia, that will be analyzed in terms of global
$m_T$-spectra which are integrated over the whole rapidity range.
In order to allow for a more direct comparison with the
$m_T$-spectra from $pN$ collisions in Figs. 5--7 we will first
discuss a more transparent situation and 'switch off' a couple of
reaction channels; the results from the 'full' calculations will
be presented in Section 3.4. To disentangle various dynamical
effects we thus first suppress the decay $\phi \rightarrow
K\bar{K}, \pi \rho$ to allow for a direct evaluation of the
$\phi$-meson $m_T$-spectra at the end of the calculation.
Furthermore, it is well known experimentally \cite{QM96,QM97,QM99}
that the apparent slope of $m_T$-spectra for different hadrons
varies almost linearly with the hadron rest mass due to a common
collective flow velocity $\beta$. In the transport calculations
this collective flow results from elastic collisions between the
hadrons in the expansion phase of the reaction
\cite{lena00,Heinz95} ('pion wind'). For our exploratory study we
switch off the elastic collisions of kaons and $\phi$-mesons in
the expansion phase with pions. This then leads to $m_T$-spectra
with roughly the same slope for all hadrons and their relative
abundance can be extracted in a simple way (see below). As mentioned before, the
related changes of the $m_T$ spectra due to elastic collisions
will be addressed in Section 3.4.

\subsection{SPS energies}

We now turn to the results of the HSD transport calculations. In
Fig. 9 we show the time evolution of $c, \bar{c}$ production
(solid line) for a central (b = 1 fm) collision of $Au + Au$ at
160 A$\cdot$GeV in comparison to $s, \bar{s}$ production (dashed
line), where the $c, \bar{c}$ number is scaled in height to the
$s, \bar{s}$ line for the initial 'hard' production by a factor of
$1.5 \times 10^3$. Both functions rise steeply within about 1
fm/c; whereas the solid line ($c,\bar{c}$) stays practically
constant the dashed line ($s, \bar{s}$) increases smoothly due to
secondary and ternary $s\bar{s}$ production by meson-baryon or
meson-meson collisions \cite{Geiss}. This 'cooking' of strangeness
in the expanding 'fireball' leads to a moderate ($\sim 46$ \%)
enhancement of strangeness whereas the secondary production of
$c\bar{c}$ pairs (by meson-baryon collisions) proceeds early
and is only $\sim 9\%$
(for the cross sections specified in Section 2), which might be
neglected at SPS energies. In this respect charm quark pairs
dominantly are created in the initial high density phase of the
collision with energy densities even above 3 GeV/fm$^3$
\cite{CassKo}. The multiplicity of open charm mesons here is about
0.2, whereas the multiplicity of $J/\Psi$'s (including the decay
of $\chi_c$) is only about $10^{-3}$. The fraction of charmonia
dissociated by baryons and mesons is 70\% for $J/\Psi$, 80 \% for
$\chi_c$ and 90 \% for $\Psi^\prime$, which is comparable to the
suppression calculated earlier in Ref. \cite{Cass97}.

The effect of rescattering of $D$-mesons on baryons and mesons as
well as charmonium interactions with hadrons is shown in Fig. 10
for a central collision of $Au + Au$ with respect to the
transverse mass spectra. Here the $D, D^*$ spectrum is flattened
out in transverse mass while the charmonium spectrum is roughly
reduced by a factor 3-4 due to dissociation reactions; the 
kink from $\Psi^\prime$ at $\sim$ 3.7 GeV disappears due to the
large $\Psi^\prime$ dissociation. We point
out that a drastic enhancement of the slope of the $D$-meson
$m_T$-spectra as advocated in Refs. \cite{Wang,Lin00} is not 
seen from our dynamical calculations for the cross sections adopted.
Quite remarkably, the open charm spectra and charmonium spectra
appear to scale well in transverse mass after including the
secondary interactions with hadrons.

The effect of final state
interactions on the rapidity distribution of $D$-mesons is
displayed in  Fig. 11 which shows a slight broadening of the
distribution due to the elastic scattering processes with baryons
and mesons, respectively.

It is interesting to have a look at the $m_T$-spectra for all
mesons in analogy to Fig. 6 (for $pp$ reactions) to explore the
effects of open charm and charmonium rescatterings. The calculated
$m_T$-spectrum for pions, kaons, $\phi$-mesons, all $D + \bar{D}$
mesons and charmonia is given in Fig. 12 which can be
characterized  well by an exponential slope parameter of 176 MeV
(dashed line) for all mesons. This result comes about as follows
when compared to Fig. 6: The $D$-mesons are created more
abundantly than pions (relative to $pp$) in central collisions of
$Au + Au$ because the $D$-meson (and charmonium) yield scales with
the number of hard collisions (cf. Fig. 8) while the pions roughly
scale with the number of participants and may be reabsorped to
some extent. The kaon (and $\phi$) yield increases due to
rescattering (cf. Fig. 9), however, the $\phi$ $(s\bar{s})$ mesons
do not match the $m_T$-scaling in the HSD transport approach and
stay down by a factor of about 3-4. The charmonium spectrum
(relative to $pp$) is decreased by a factor $\approx$ 3-4 due to
dissociation processes as noted before. All these effects lead to
the approximate $m_T$-scaling without employing the assumption of
a Quark-Gluon Plasma (QGP) formation and a common hadronization at
some temperature of 160 -- 180 MeV.

We mention that a roughly constant $\pi$ to $J/\Psi$ ratio from
$pp$ to central $Pb + Pb$ collisions has been also pointed out in
Refs. \cite{Marek1,Marek2} proposing a statistical hadronization
scheme in all reactions. Furthermore, Gallmeister {\it et al.}
have suggested in Ref. \cite{Gall} that the open charm degrees of
freedom might be described in a simple thermodynamical model for
central collisions of $Pb + Pb$ at the SPS using the same
temperature for all mesons. The findings of these authors are
supported here by the nonequilibrium transport calculations, that
provide rather simple arguments for the phenomena pointed out
before.

\subsection{$Au + Au$ reactions at 25 A$\cdot$GeV}

In this Section we explore the perspectives of open charm
measurements in nucleus-nucleus collisions at 25 A$\cdot$GeV,
which might be accessable at a possible future GSI facility
\cite{GSI}. In this initial study we restrict to central
collisions of $Au + Au$ at 25 A$\cdot$GeV, which are expected to
provide the optimal conditions for open charm experiments and
studies on the in-medium properties of $D$-mesons in analogy to
the $K^+, K^-$ experiments at the SIS. We step ahead as in Section
3.1.

Fig. 13 shows the time evolution of $c, \bar{c}$ production (solid
line) for a central (b = 1 fm) collision in comparison to $s,
\bar{s}$ production (dashed line) where the number of $c, \bar{c}$
is scaled again in height to the $s, \bar{s}$ line for the initial
'hard' production (by a factor 1.5$\times 10^5$). Both functions
rise within a few fm/c which corresponds to the passage time of
the (Lorentz contracted) nuclei. As in Fig. 9 the solid line
($c,\bar{c}$) stays  constant for later times while the dashed
line ($s, \bar{s}$) increases again due to secondary and ternary
$s\bar{s}$ production channels. The relative enhancement of
$s\bar{s}$ 'cooking' here amounts to roughly 65\% whereas the
relative contribution of $c\bar{c}$ pairs from secondary channels
is $\sim 7 \%$ for the cross sections specified in Section 2. We
note, however, that the 65\% enhancement of strangeness is
insufficient to explain the $K^+$ abundancies at the AGS
\cite{AGSK} from 4 - 11 A$\cdot$GeV or the $K/\pi$ ratio at 40 A
GeV (at the SPS) without assuming any in-medium modifications of
the kaons. For a detailed discussion we refer the reader to Refs.
\cite{Cass00,Geiss}.

It is apparent from Fig. 13 that the charm quark pairs are created
in the initial high density phase of the collision, here with
energy densities up to 2 GeV/fm$^3$, which is above the critical
energy density from lattice calculations for the formation of a
QGP  \cite{lattice}. However, the energy densities from the
transport calculation correspond to nonequilibrium phase-space
configurations at high baryon density, that should not be
identified with the energy density extracted from lattice
calculations (in equilibrium and for quark chemical potential $\mu_q = 0$).

For a quantitative orientation we display in Fig. 14 the volume
(in the nucleus-nucleus center-of-mass) with an energy density
above 1 GeV/fm$^3$ and 2 GeV/fm$^3$ as a function of time for a
central $Au + Au$ collision at 25 A$\cdot$GeV, where only
interacting and produced hadrons have been counted as in Ref.
\cite{CassKo}. It is
important to note that the high energy density is essentially
build up from 'strings', i.e. 'unformed' hadrons. This phase may
be addressed as {\it string matter} (cf. Refs.
\cite{Pradip,Weber}) and expresses the notion that most of the
hadrons appear in some form of 'continuum excitation'. The energy
density including only 'formed' hadrons (during the expansion of
the system) stays below 1 GeV/fm$^3$, i.e. below the energy
density expected for a transition to the QGP. The absolute numbers
in Fig. 14 have to be compared to the volume of a $Au$-nucleus in
the moving frame which, for a Lorentz $\gamma$-factor of 3.78,
gives $\approx $ 330 fm$^3$. Thus also at 25 A$\cdot$GeV the phase
boundary to a QGP might be probed in a sizeable volume for time
scales of a few fm/c. Contrary to central collisions at the SPS
these volumes are characterized by a high net quark density; for
such configurations we presently have no reliable guide from QCD
lattice calculations.

The multiplicity of open charm mesons at 25 A$\cdot$GeV is about
$6\cdot 10^{-4}$, whereas the multiplicity of $J/\Psi$'s
(including the decay of $\chi_c$) is about $1.5 \cdot 10^{-5}$. We
mention that the fraction of charmonia dissociated by baryons and
mesons is $\sim$ 60\% for $J/\Psi$ \cite{Cass00}.

The effect of rescattering of $D$-mesons on baryons and mesons is
displayed in Fig.  15 (for a central collision of $Au + Au$) for the
$D$-meson rapidity distribution, which shows now a substantial
broadening due to scattering processes with baryons and mesons. The
decrease of the $D$-meson rapidity distribution at midrapidity is
almost a factor of 2.

The $m_T$-spectra for all mesons in analogy to Fig. 5 (for $pp$
reactions) are presented in Fig. 16. The calculated $m_T$-spectrum
for pions, kaons and $\phi$-mesons can be characterized  by an
exponential slope parameter of 143 MeV (dashed line). Again the
kaon (and $\phi$) yield is increased (relative to $pp$ times the
number of hard collisions $N_{coll}$) due to rescattering (cf.
Fig. 13), but the $\phi$ $(s\bar{s})$ mesons stay down by a factor
of  3-4. The charmonium spectrum (relative to $pp$) is decreased
by a factor $\approx$ 2.5 due to dissociation as noted before and
approximately fulfills the global $m_T$-scaling.  The latter does
not hold for $D$-mesons (open squares) which are suppressed
dynamically in the threshold region by roughly one order of
magnitude.

We recall that a similar observation has been made for the
$m_T$-scaling of $K^+$ and $K^-$ mesons close to threshold
energies at the SIS \cite{Elena}, where the strange mesons have
been suppressed relative to pions and $\eta$'s. However, when
adding to the $K^+$ mass the $\Lambda-N$ mass difference of 177
MeV (due to the associated production mechanism in $pp$ and $\pi
N$ collisions), a remarkable $m_T$-scaling could be recovered
again \cite{Elena}. It should be noted that the latter scaling is
not due to a grand-canonical (or canonical)  chemical equilibration, but simply
due to a shift of the spectra induced by the kaon production
mechanism. We have to stress, however, that all these observations
on the charm sector are based on our extrapolations (Section 2)
and might not hold experimentally.

We now address the question, to what extent in-medium
modifications of the $D$-mesons might be seen  in the
$m_T$-spectra for central $Au+Au$ collisions at 25 A GeV. Contrary
to open charm production and propagation in antiproton induced
reactions on nuclei \cite{Alex99}, where the $D$-mesons show up
with momenta of a couple of GeV/c relative to the nuclear matter
rest frame, the $D$-mesons produced in central nucleus-nucleus
collisions have only small momenta in the rest frame of the
hadronic fireball. This is of particular relevance for
experimental studies of hadron self energies, since the latter are
generally momentum dependent and most pronounced for low momenta.

The modifications of the $D$-meson spectral functions in the
medium at present cannot be reliably calculated nor extracted (in
the low density limit) from experimental scattering data via a
dispersion analysis (cf. Ref. \cite{Sib98} for the $K,\bar{K}$
problem). For our initial study we thus discard all momentum
dependence of the $D$-meson self energies and also neglect a
broadening of their spectral functions due to interactions in the
medium \cite{offshell}. As a guide we employ the QCD sum rule
calculations from Ref. \cite{Weise} and implement a mass shift of
the form
\begin{equation}
 \Delta m_D (\rho) = \alpha_D \frac{\rho}{\rho_0}
\label{drop}
\end{equation}
with $\alpha_D \approx - 50$ MeV, where  $\rho_0$ denotes the
nuclear matter density and $\rho$ the actual baryon density at the
$D$-meson creation point. In principle one might expect different
mass shifts of $D$ and $\bar{D}$ mesons in the medium due to an
opposite sign of the vector interaction \cite{Alex99}. However,
since the $c, \bar{c}$ pairs are created in the early high density
phase of the collision (cf. Fig. 13) the vector interaction is
expected to vanish here (cf. Refs. \cite{Pradip,Gerry}) and only
the scalar attraction to survive. This will lead to similar mass
shifts for $D$ and $\bar{D}$ mesons as anticipated in Eq.
(\ref{drop}). Since densities up to $8 \rho_0$ can be achieved in
central $Au+Au$ collisions at 25 A GeV, the $D$-meson mass shifts
may reach up to -400 MeV. Such mass shifts have a dramatic effect
on the production cross sections in $pN$ collisions and secondary
meson-baryon reactions when incorporating them in the production
thresholds (cf. Tables 1 and 2).

Our calculations with the mass shift (\ref{drop}) (crosses in Fig. 16) give an
enhancement of the $D$-meson yield by about a factor of 7 relative
to the bare-mass case (open squares). The slope of the spectra is
modified only slightly relative to the bare mass case as can
be seen from Fig. 16 for the resulting $m_T$ spectrum.
Somewhat surprisingly, an approximate $m_T$-scaling with all other
mesons is regained in this case. We have to point out again that
the results on open charm and charmonia in Fig. 16 essentially
depend on our extrapolations in Section 2 and the assumed self
energies (\ref{drop}), which are not controlled by data. On the
other hand, Fig. 16 should be helpful in guiding the experimental
analysis.

\subsection{Central collisions of $Au + Au$ at $\sqrt{s}$ = 200
GeV} Apart from the low energy (threshold) regime, explored in
Section 3.2, we also present predictions for central $Au + Au$
collisions at RHIC energies of $\sqrt{s}$ = 200 GeV or 21.3
A$\cdot$TeV, which will be investigated in the near future. For an
overview of the predictions performed within the HSD approach we
refer the reader to Ref. \cite{Cass00}. Here we extend our
calculations to open charm mesons and provide transverse mass
spectra for all mesons in analogy to Figs. 12 and 16.

The multiplicity of open charm pairs for $Au + Au$ at $b$=1 fm and
 21.3 A$\cdot$TeV from the HSD approach is about
$16$, whereas the multiplicity of final $J/\Psi$'s (including the
decay of $\chi_c$) is only  $\sim 8 \cdot 10^{-2}$ since  in this
case the dissociation by baryons and mesons is $\sim$ 90\% for
$J/\Psi$ \cite{Cass00}. We mention that the production of charm
pairs proceeds within less than 0.8 fm/c (in the nucleus-nucleus
cms) and that the amount of $c\bar{c}$-pairs from secondary
meson-baryon reactions is approximately 11 \% for the cross
sections specified in Section 2.

The $m_T$-spectra for all pions, kaons and antikaons,
$\phi$-mesons, $D$ and $\bar{D}$ mesons as well as charmonium
states  are presented in Fig. 17 in analogy to Fig. 7 (for $pp$
reactions). The calculated $m_T$ spectrum for pions, kaons,
$\phi$-mesons can be characterized  by an exponential slope
parameter of 225 MeV (dashed line). Similar to the lower
bombarding energies in Sections 3.1 and 3.2 the kaon (and $\phi$)
yield is increased  due to rescattering in the hot and initially
high density mesonic fireball. The charmonium spectrum (relative
to $pp$) is decreased by a factor $\approx$ 10 due to dissociation
as noted before and no longer fulfills the global $m_T$-scaling.
The  $D$-mesons (open squares) are somewhat enhanced relative to
the $m_T$-scaling which, however, might be an artefact of the
parametrizations in Section 2.  On the other hand, the relative
suppression of charmonia could be compensated by $D\bar{D}$ flavor
exchange reactions to $J/\Psi + \pi$ etc., i.e. the inverse
channels responsible for charmonium dissociation in interactions
with mesons \cite{Rafelski}. However, as mentioned before, these
cross sections are presently not sufficiently known such that a
final answer on the relative importance of
these subsequent charmonium production channels has to wait for
future.

\subsection{Collective acceleration of mesons in the expansion
phase}
 As mentioned above, the $m_T$-spectra from Figs. 12, 16 and 17 provide a
 global view on the effect of chemical (inelastic) reactions, but should
not be compared with experiment directly since the rescattering of
kaons and $\phi$ mesons with pions has been switched off. In order
to demonstrate the effect of the elastic scatterings, that have
been discarded in Sections 3.1 - 3.3, we present in Fig. 18 the
results for all $m_T$ spectra from the transport calculations for
$Au + Au$ at 25 A$\cdot$GeV, 160 A$\cdot$GeV and 21.3 A$\cdot$TeV
now including the elastic rescatterings. Here the slopes of the
pions are slightly decreased whereas the slopes for kaons and
$\phi$-mesons increase at all bombarding energies. The spectra for
the open charm mesons as well as charmonia do not change within
the numerical accuracy since their rescattering with baryons and
mesons had already been taken into account in the previous
calculations in Sections 3.1 -- 3.3. We note explicitly, that the
high slope parameter for $\phi$-mesons of $\sim$ 300 MeV seen
experimentally at midrapidity in central collisions of $Pb+Pb$ at
the SPS by NA49 \cite{Stock} is not reproduced within the HSD
calculations due to the weak coupling of $\phi$-mesons to
non-strange hadrons. If this phenomenon is related to an early
acceleration of strange quarks and antiquarks in a QGP phase or
due to unexpected large rescattering cross sections is presently
unclear.

For comparison we discuss a scenario where the $D,\bar{D}$ mesons
and charmonia emerge from a hadronizing QGP at rather low hadron
density as advocated in \cite{Johanna}. In the latter scenario the
collective expansion with velocity $v_{\perp}$ results from the
pressure in the QGP phase and all hadrons freeze out at the same
quark chemical potentials $\mu_q, \mu_s, \mu_c$ and temperature
$T_{therm}$. The slopes of open charm mesons and charmonia then
are expected to change as $T_X \approx T_{therm} + m_X/2
v_{\perp}^2$ for low momenta and as $T_X \approx T_{therm}
\sqrt{(1+v_{\perp})/(1-v_{\perp})}$ for momenta $p_{\perp} \gg
m_X$, where $v_{\perp} = \beta$ is a collective velocity in the
range 0.4 $\le v_{\perp} \le$ 0.6. Thus $D,\bar{D}$ mesons and
charmonia ($J/\Psi, \Psi^\prime$) should show effective slopes
larger than 0.35 GeV at SPS
and RHIC energies. This conjecture might be tested soon
experimentally and prove or disprove the moderate slopes of open
charm mesons as predicted within the hadron-string-dynamics (HSD)
approach.

\subsection{Excitation functions of mesons in central collisions}

In order to provide a more complete overview on meson production
we show in Fig. 19 the excitation function of open charm mesons in
central $Au + Au$ collisions from AGS to RHIC energies without
employing any self energies for these mesons. The $\bar{D}$-mesons
with a $\bar{c}$ are produced more frequently at low energies due
to the associated production with $\Lambda_c, \Sigma_c,
\Sigma_C^*$ similar to the kaon case (cf. lower part). At roughly
15 A$\cdot$GeV the cross sections for open charm and charmonia are
similar, while the ratio of open charm to charmonium bound states
increases rapidly with energy. This behaviour is quite similar to
the excitation functions in the strangeness sector when comparing
$K^+,K^-$ and $\phi$-mesons. Since the excitation function for
open charm drops very fast with decreasing bombarding energy,
experiments around 20 A$\cdot$GeV will be a challenging task since
the multiplicity of the other mesons is higher by orders of
magnitude. On the other hand, the perspectives for open charm
measurements at RHIC appear promising since about 16 $c\bar{c}$
(or $ D\bar{D})$ pairs should be created in central $Au + Au$
collisions according to our calculations.

We mention that the excitation functions for the pions, kaons,
eta's and $\phi$-mesons have been taken from Ref. \cite{Cass00},
while the multiplicities for $J/\Psi$ have been recalculated using
the novel comover absorption cross sections from Section 2 and
Ref. \cite{Haglin} as well the parametrization (\ref{fitj}) instead
of the Schuler fit \cite{schuler}. Since the numbers up to 500 A$\cdot$GeV
are compatible within 30\%
we do not discuss these differences in more detail. The higher
$J/\Psi$ multiplicity at RHIC energies is a direct consequence of
the cross section (\ref{fitj}).

\section{Summary}

In this work we have calculated excitation functions for open
charm mesons in central $Au+Au$ reactions from AGS to RHIC
energies within the HSD transport approach. The 'input' open charm
cross sections from $pp$ and $\pi N$ reactions have been fitted to
results from PYTHIA and scaled in magnitude to the available
experimental data. In order to study the relative changes from
central $Au + Au$ to $pp$ collisions, we have first switched off
elastic final state interactions of kaons and $\phi$-mesons with
pions in order to suppress their common acceleration in the 'pion
wind' during the expansion phase. Within the parametrizations and
results from the LUND string model \cite{LUND} -- which is
incorporated in the HSD approach -- we find an $m_T$-scaling for
pions, kaons, $D$-mesons and $J/\Psi$ in central collisions of $Au
+ Au$ at the SPS (with an apparent slope of 176 MeV) without
employing the assumption of a Quark-Gluon Plasma (QGP) and common
freeze-out properties. We have shown that this result is
essentially due to an approximate $m_T$-scaling in $pp$ collisions
at $\sqrt{s}$ = 17.3 GeV and $D, \bar{D}$ and $J/\Psi$ final state
interactions. Furthermore, final state elastic scatterings change
this conclusion to a moderate extent since the relative meson
abundancies are not altered anymore and their spectra only get
modified due to a common collective acceleration.

At bombarding energies of 25 A$\cdot$GeV a suppression of
$D$-mesons by a factor of $\sim$ 10 relative to a global
$m_T$-scaling with a slope of 143 MeV is expected if no $D$-meson
self energies are accounted for. On the other hand, attractive
mass shifts of -50 MeV at $\rho_0$ -- when extrapolated linearly
in the baryon density -- lead to an enhancement of open charm
mesons by about a factor of 7 such that an approximate
$m_T$-scaling for all mesons (cf. Fig. 16) is regained.

At RHIC energies of $\sqrt{s}$ = 200 GeV or 21.3 A$\cdot$TeV the
global $m_T$-scaling is expected to hold also within a factor of
2-3 except for the charmonium states which -- within the HSD
transport approach -- are dissociated by baryons and 'late
comovers' to $\sim 90 \%$. On the other hand, the inverse reaction
channels $D+\bar{D} \rightarrow J/\Psi +$ meson etc. might lead
again to charmonium enhancement as suggested in Ref.
\cite{Rafelski}. Here we leave this question open for future
analysis.

However, as pointed out throughout this work, the elementary cross
sections for open charm and charmonia in $pp$ and $\pi N$
reactions have to be measured in the relevant kinematical regimes
before reliable conclusions can be drawn in the nucleus-nucleus
case. Experimental data in the 20 - 30 A$\cdot$GeV with light and
heavy systems will have to clarify, furthermore, if the
quasi-particle picture of open charm mesons at high baryon density
is applicable at all.

\acknowledgements
The authors are grateful to C. Greiner and R. Vogt for helpful
discussions and to P. Senger for valuable suggestions.


\begin{table}[t]
\begin{center}
\caption{The parameters $a_x, \alpha$ and $\beta$ for $pN$
reactions
\label{tabl1} }
\vspace{0.5cm}
\begin{tabular}{ |lllll| }
\hline
\multicolumn{5}{|c|}{$pN$} \\ \hline
 Meson         & $\sqrt{s_0}$ [GeV] & $a_x$ [mb] & $\alpha$ & $\beta$
\\ \hline
$D^0$         & 5.605 & 0.523 & 4.92 & 1.36 \\
$\bar D^0$    & 5.069 & 0.496 & 4.96 & 1.36 \\
$D^+$         & 5.609 & 0.469 & 4.76 & 1.40 \\
$D^-$         & 5.073 & 0.363 & 4.94 & 1.44 \\ \hline
$D^{0*}$      & 5.889 & 1.775 & 4.90 & 1.34 \\
$\bar D^{0*}$ & 5.230 & 1.275 & 4.56 & 1.42 \\
$D^{+*}$      & 5.896 & 1.514 & 4.64 & 1.40 \\
$D^{-*}$      & 5.233 & 1.384 & 5.20 & 1.36 \\ \hline
$D^+_s$       & 5.813 & 0.171 & 5.12 & 1.34 \\
$D^-_s$       & 5.373 & 0.102 & 5.58 & 1.42 \\
$D^{+*}_s$    & 6.101 & 0.496 & 4.88 & 1.38 \\
$D^{-*}_s$    & 5.516 & 0.283 & 5.50 & 1.46 \\ \hline
\end{tabular}
\end{center}
\end{table}

\begin{table}[h]
\begin{center}
\caption{ The parameters $a_x, \alpha$ and $\beta$ for $\pi N$
reactions
\label{tabl2} }
\vspace{0.5cm}
\begin{tabular}{ |lllll| }
\hline
\multicolumn{5}{|c|}{$\pi N$} \\ \hline
 Meson         & $\sqrt{s_0}$ [GeV] & $a_x$ [mb] & $\alpha$ & $\beta$
\\ \hline
$D^0$         & 4.667 & 0.273 & 2.86 & 1.28 \\
$\bar D^0$    & 4.150 & 0.247 & 3.80 & 1.26 \\
$D^+$         & 4.671 & 0.255 & 3.22 & 1.28 \\
$D^-$         & 4.154 & 0.286 & 3.50 & 1.22 \\ \hline
$D^{0*}$      & 4.951 & 1.076 & 3.14 & 1.22 \\
$\bar D^{0*}$ & 4.292 & 0.774 & 3.80 & 1.26 \\
$D^{+*}$      & 4.955 & 0.719 & 2.86 & 1.32 \\
$D^{-*}$      & 4.296 & 0.839 & 3.40 & 1.24 \\ \hline
$D^+_s$       & 4.875 & 0.0932& 3.62 & 1.22 \\
$D^-_s$       & 4.435 & 0.0545& 3.70 & 1.34 \\
$D^{+*}_s$    & 5.162 & 0.284 & 3.42 & 1.24 \\
$D^{-*}_s$    & 4.578 & 0.163 & 3.64 & 1.34 \\ \hline
\end{tabular}
\end{center}
\end{table}

\begin{figure}[h]
\centerline{\psfig{file=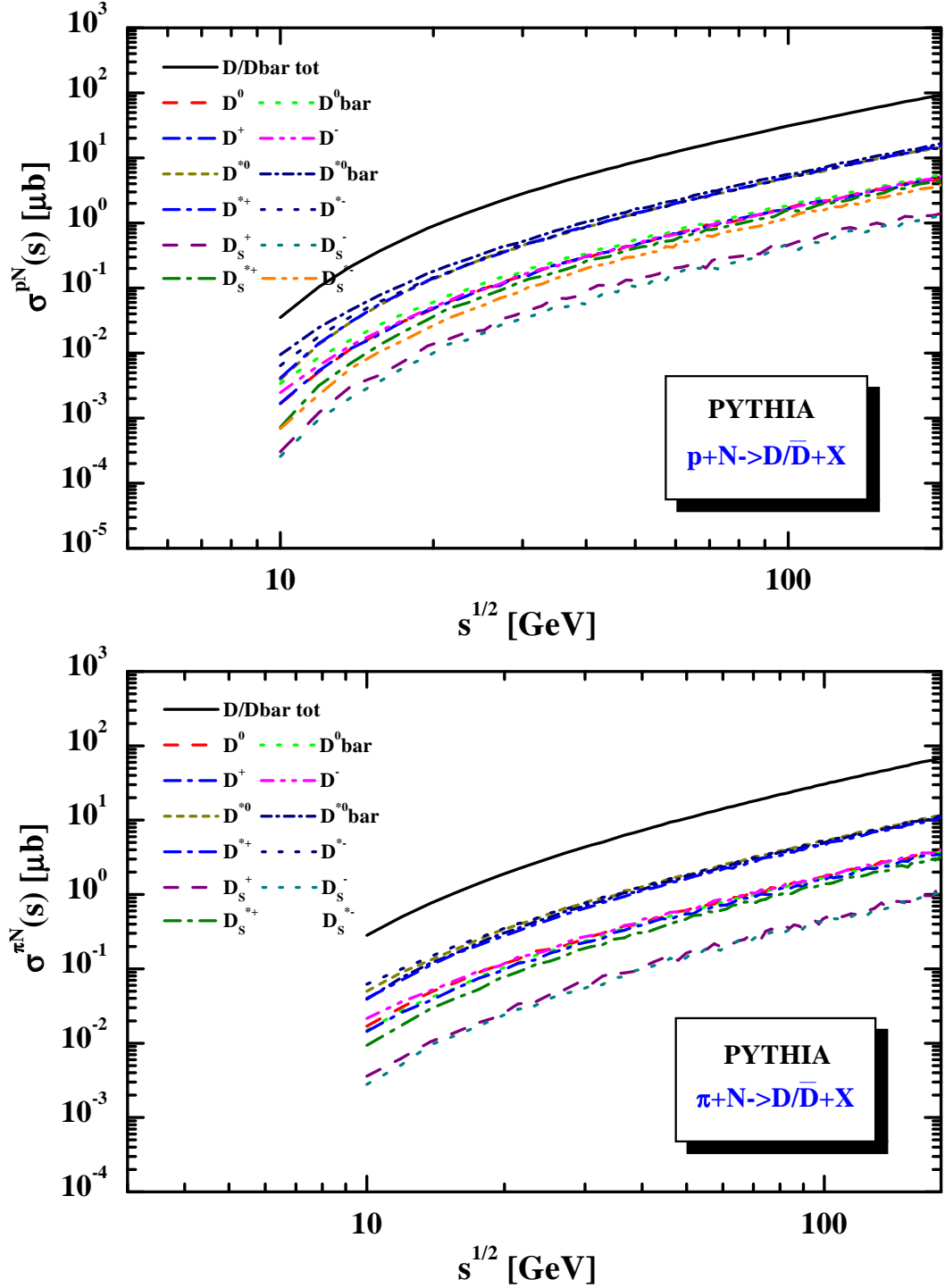,width=14cm}}
\vspace*{5mm}
\caption{The cross section for open charm mesons from PYTHIA
\protect\cite{PYTHIA} for $pp$ (upper part) and $\pi N$ reactions
(lower part) using MRS G structure functions, $m_c$ = 1.5 GeV and $k_T$ =
1 GeV, respectively. The upper solid lines denote the sum over all
$D+\bar{D}$ mesons. }
\label{bild1}
\end{figure}

\begin{figure}[h]
\centerline{\psfig{file=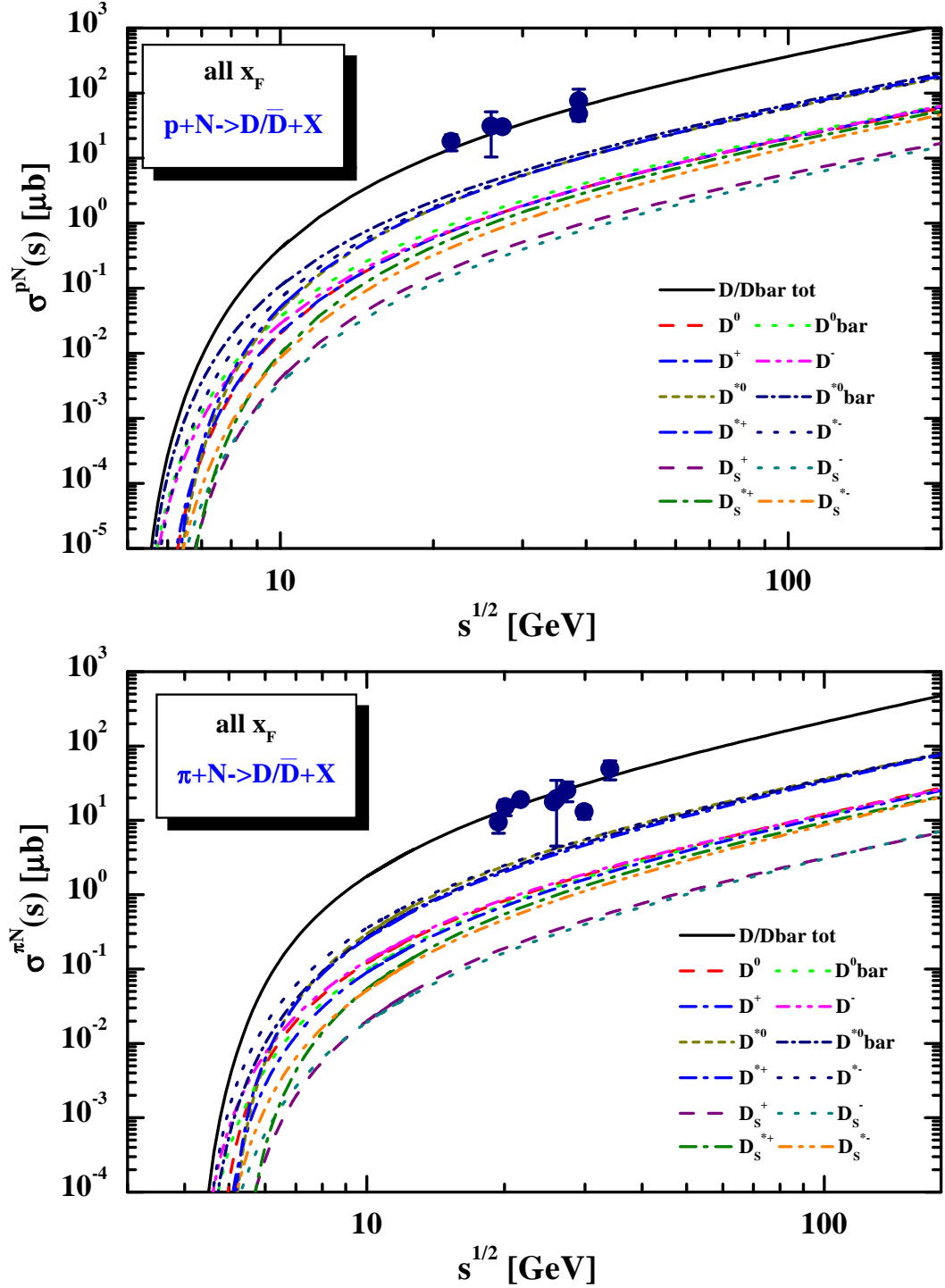,width=14cm}}
\vspace*{5mm}
\caption{The cross section for open charm mesons in the parametrization
(1) using the parameters from Tables 1 and 2 in comparison to the
experimental data from Refs. \protect\cite{NA16}-\protect\cite{E791} for $pp$ (upper part)
and $\pi N$ reactions (lower part). The upper solid lines denote the sum over
all $D+\bar{D}$ mesons.}
\label{bild2}
\end{figure}

\begin{figure}[h]
\centerline{\psfig{file=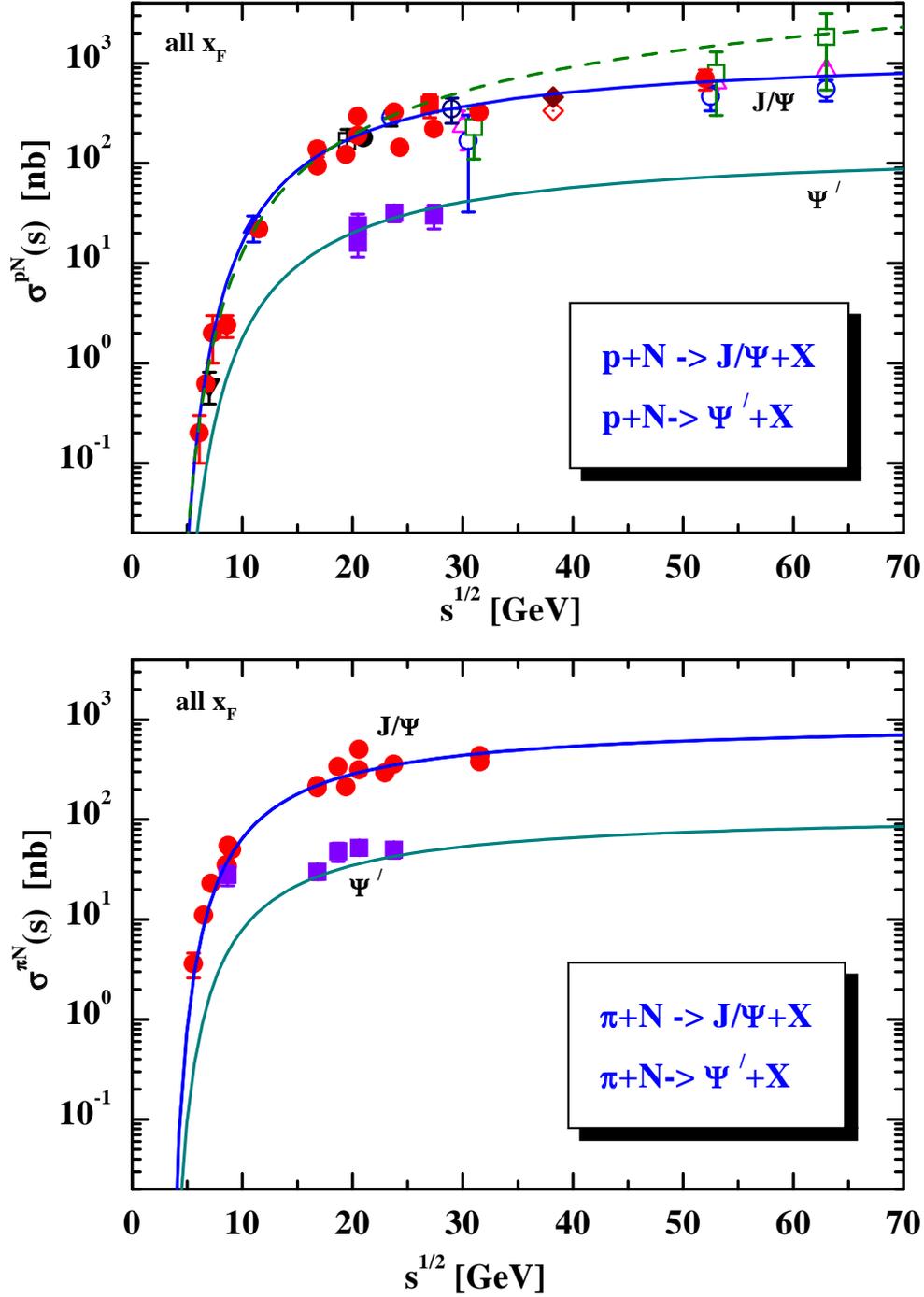,width=13cm}}
\vspace*{5mm}
\caption{The cross section for $J/\Psi$ and $\Psi^\prime$  mesons in
the parametrizations from Ref. \protect\cite{schuler} (solid lines) in
comparison to the experimental data  for $pN$ (upper part) and $\pi N$
reactions (lower part). The $J/\Psi$ cross sections include the 
decay from $\chi_c$ mesons. The dashed line in the upper part shows 
the $J/\Psi$ cross section
for the parametrization (\protect\ref{fitj}). }
\label{bild3}
\end{figure}

\newpage
\begin{figure}[h]
\centerline{\psfig{file=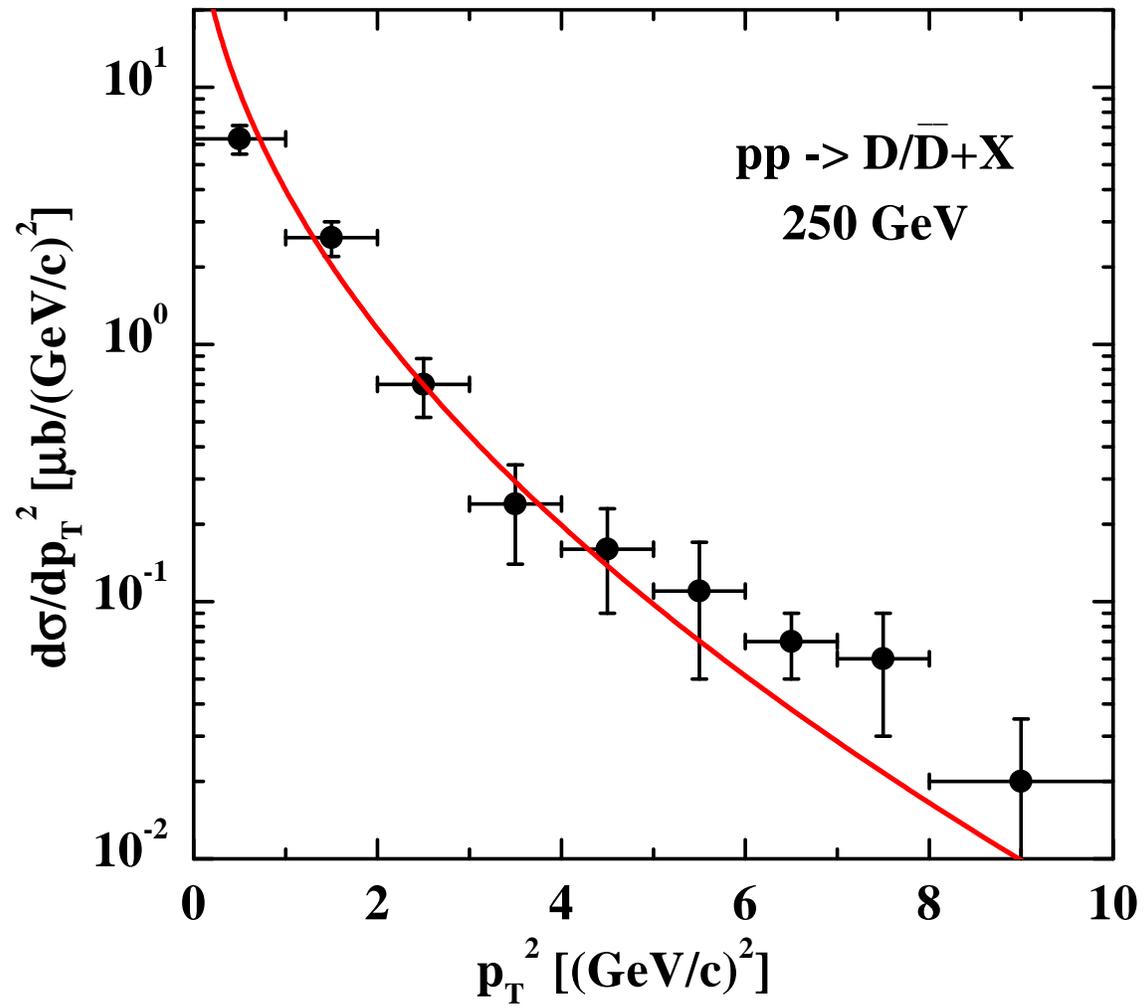,width=15cm}}
\vspace*{5mm}
\caption{The differential cross section for $D/\bar{D}$ mesons in
transverse momentum (squared) for $pp$ reactions at 250 GeV within the
parametrisation (\protect\ref{fit2}) (solid line) in comparison to the 
data from Ref. \protect\cite{E769}.}
\label{bild4}
\end{figure}

\newpage
\begin{figure}[h]
\centerline{\psfig{file=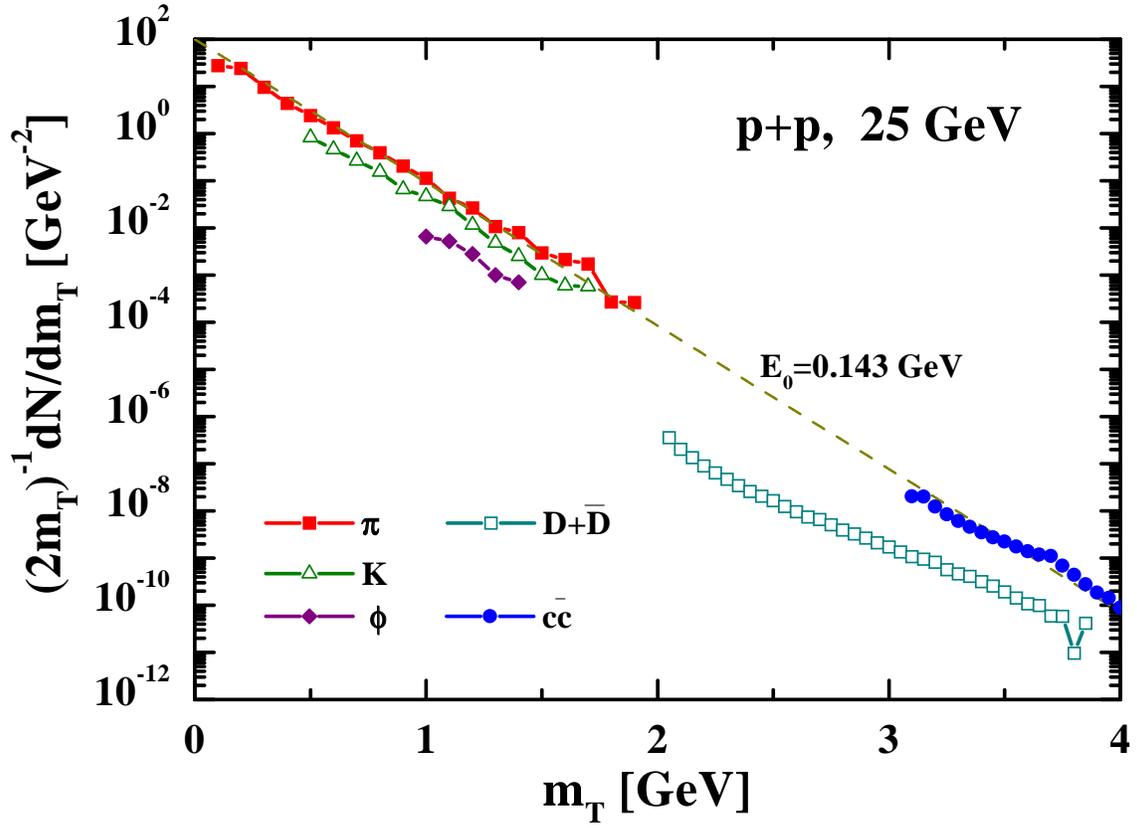,width=15 cm}}
\vspace*{5mm}
\caption{The transverse mass specta from $pp$ collisions at $T_{lab}$ =
25 GeV for pions (full squares), kaons (open triangles), and
$\phi$-mesons (full rhombes) from the LUND string model \protect\cite{LUND}
as implemented in HSD. The $D+\bar{D}$ meson (open squares) and charmonium
(full dots) spectra -- including the decay $\chi_c \rightarrow J/\Psi +
\gamma$ -- result from the parametrizations specified in Section 2. The
dashed line shows an exponential with slope parameter $E_0$ = 0.143
GeV.}
\label{bild5}
\end{figure}

\newpage
\begin{figure}[h]
\centerline{\psfig{file=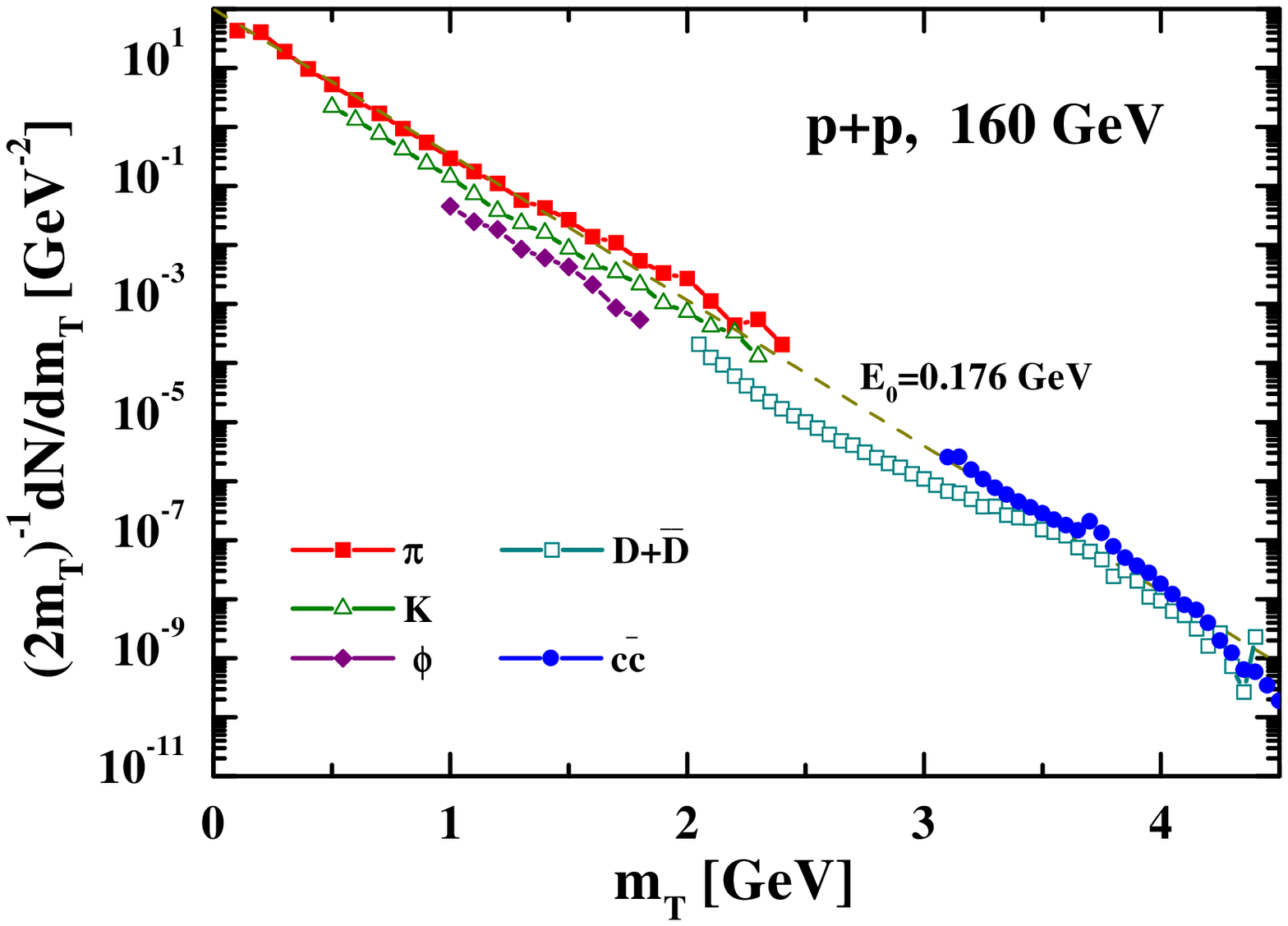,width=15 cm}}
\vspace*{5mm}
\caption{Same as Fig. 5 for $pp$ reactions at
$T_{lab}$ = 160 GeV. The dashed line shows an
exponential with slope parameter $E_0$ = 0.176 GeV.}
\label{bild6}
\end{figure}

\begin{figure}[h]
\centerline{\psfig{file=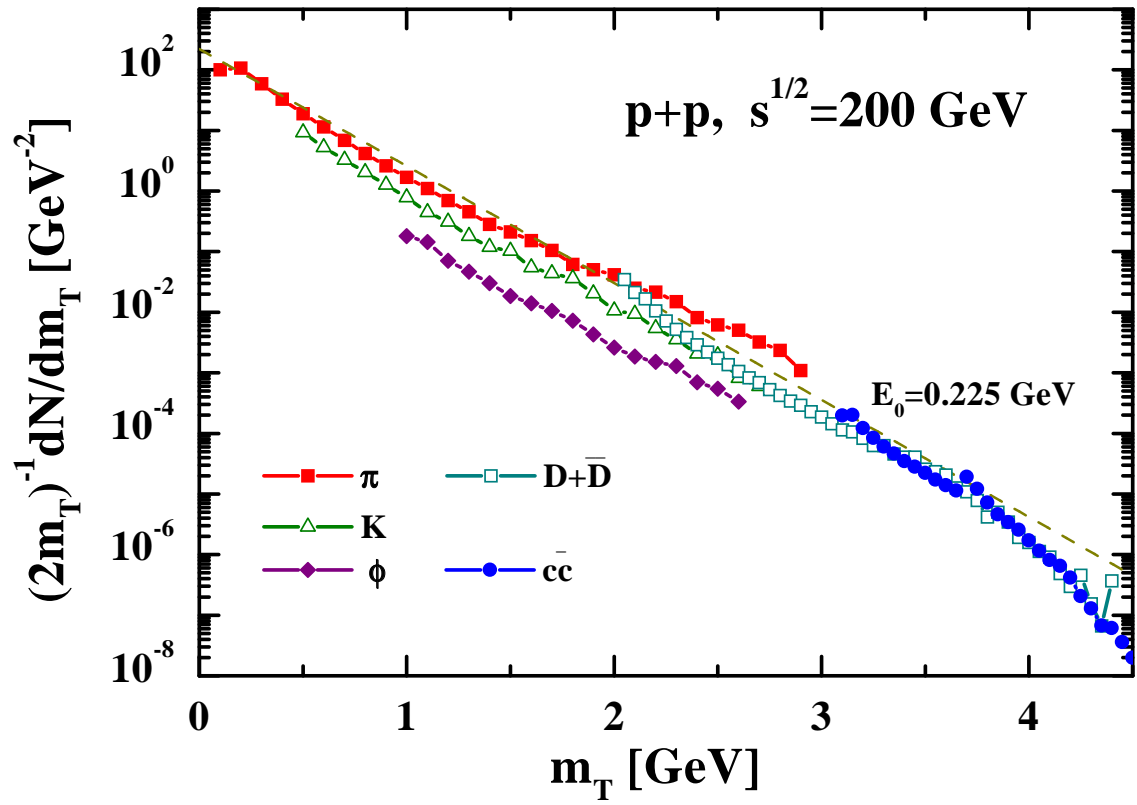,width=15 cm}}
\vspace*{5mm}
\caption{Same as Fig. 5 for $pp$ reactions at $\sqrt{s}$ = 200 GeV. The
dashed line shows an exponential with slope parameter $E_0$ = 0.225
GeV.}
\label{bild7}
\end{figure}

\begin{figure}[h]
\centerline{\psfig{file=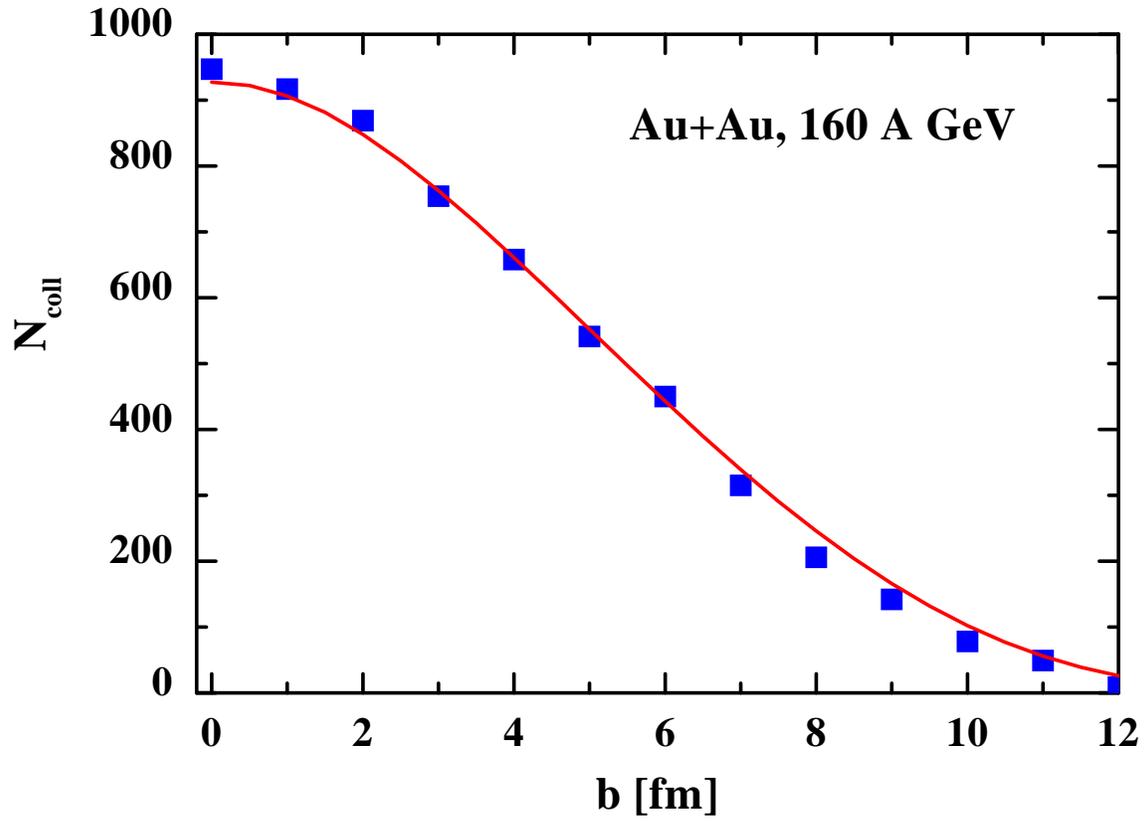,width=15 cm}}
\vspace*{5mm}
\caption{The number of hard collisions $N_{coll}$ as a function of
impact parameter $b$ in the HSD approach (full squares) for $Au + Au$
at 160 A$\cdot$GeV (see text) in comparison to the number of collisions in
the Glauber approach (solid line).}
\label{bild8}
\end{figure}

\begin{figure}[h]
\centerline{\psfig{file=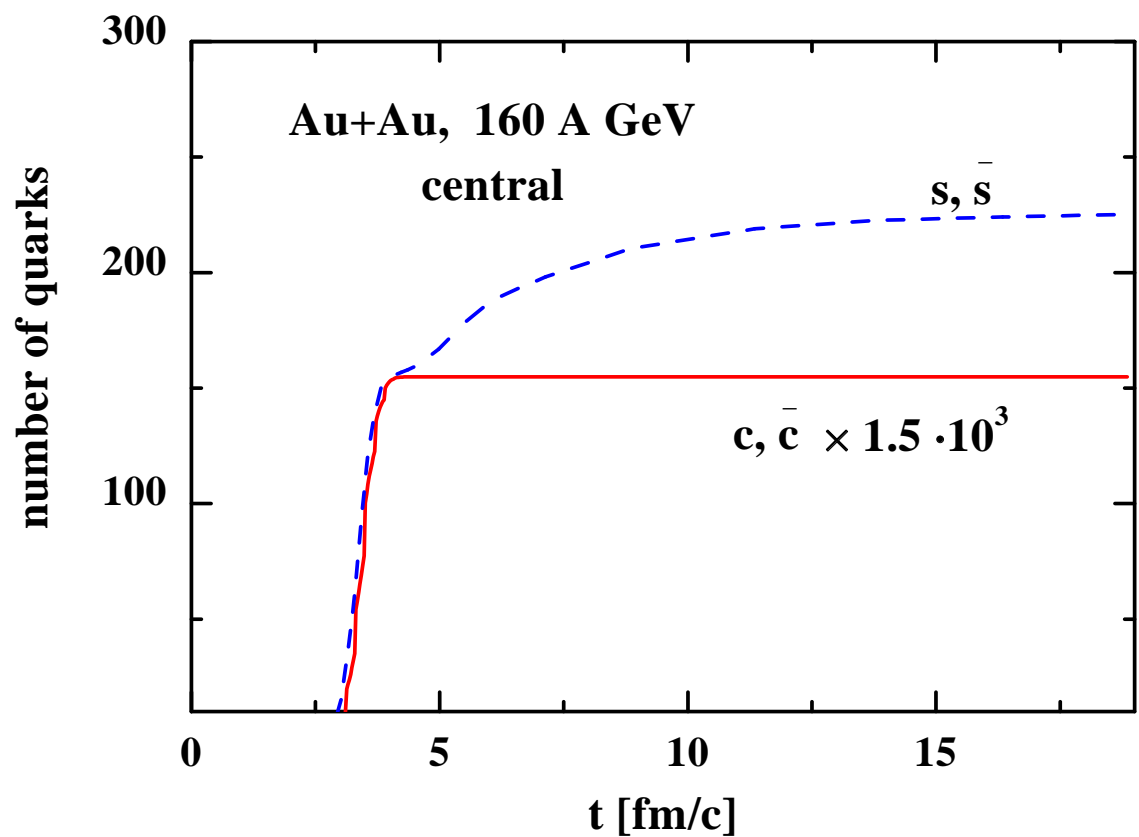,width=15 cm}}
\vspace*{5mm}
\caption{Time evolution for the production of $s, \bar{s}$ (dashed
line) and $c \bar{c}$ quarks (solid line, multiplied by a factor of $1.5 \times 10^3$)
in the HSD approach for a
central $Au+Au$ reaction at 160 A$\cdot$GeV.  The $c, \bar{c}$ numbers have been
scaled to the initial hard scattering processes.}
\label{bild9}
\end{figure}

\begin{figure}[h]
\centerline{\psfig{file=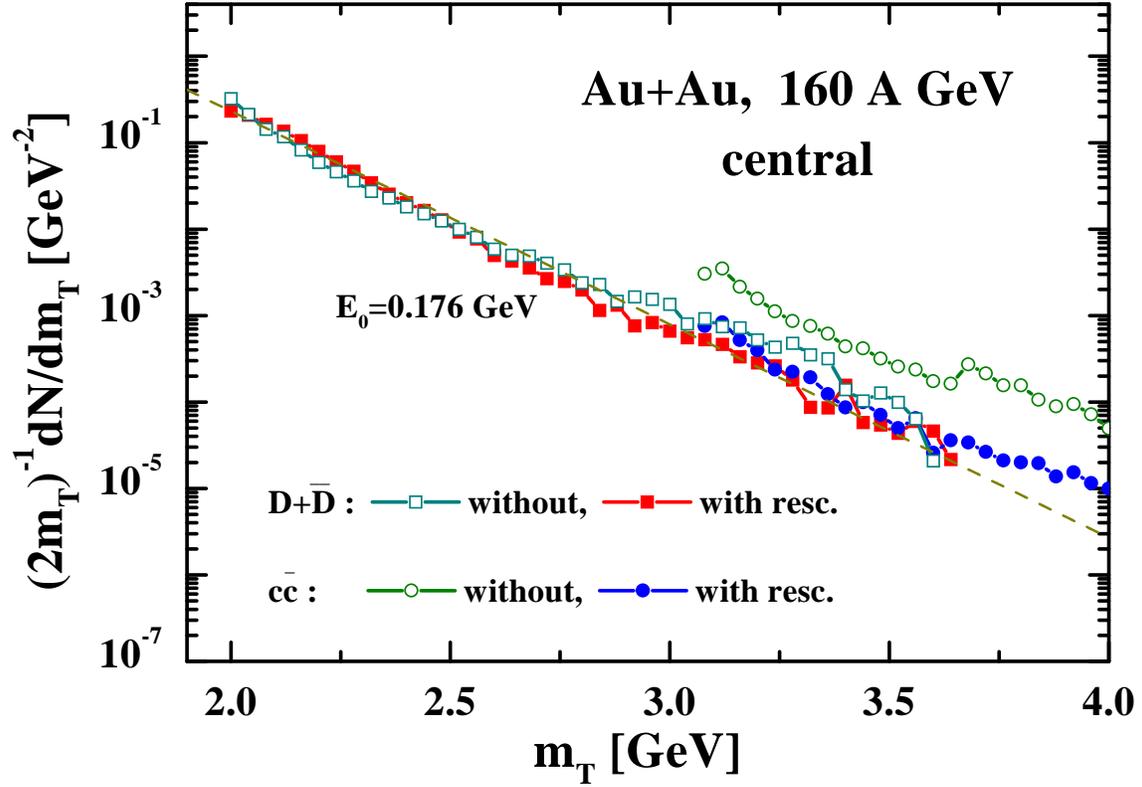,width=15cm}}
\vspace*{5mm}
\caption{The transverse mass spectra of $D+\bar{D}$ mesons and $J/\Psi,
\Psi^\prime$ mesons in the HSD approach for a central $Au+Au$ collision
at 160 A$\cdot$GeV.  The open symbols denote the spectra without rescattering
and reabsorption while the full symbols include the final state
interactions. The thin dashed line shows an exponential with slope
parameter $E_0$ = 0.176 GeV.}
\label{bild10}
\end{figure}

\begin{figure}[h]
\centerline{\psfig{file=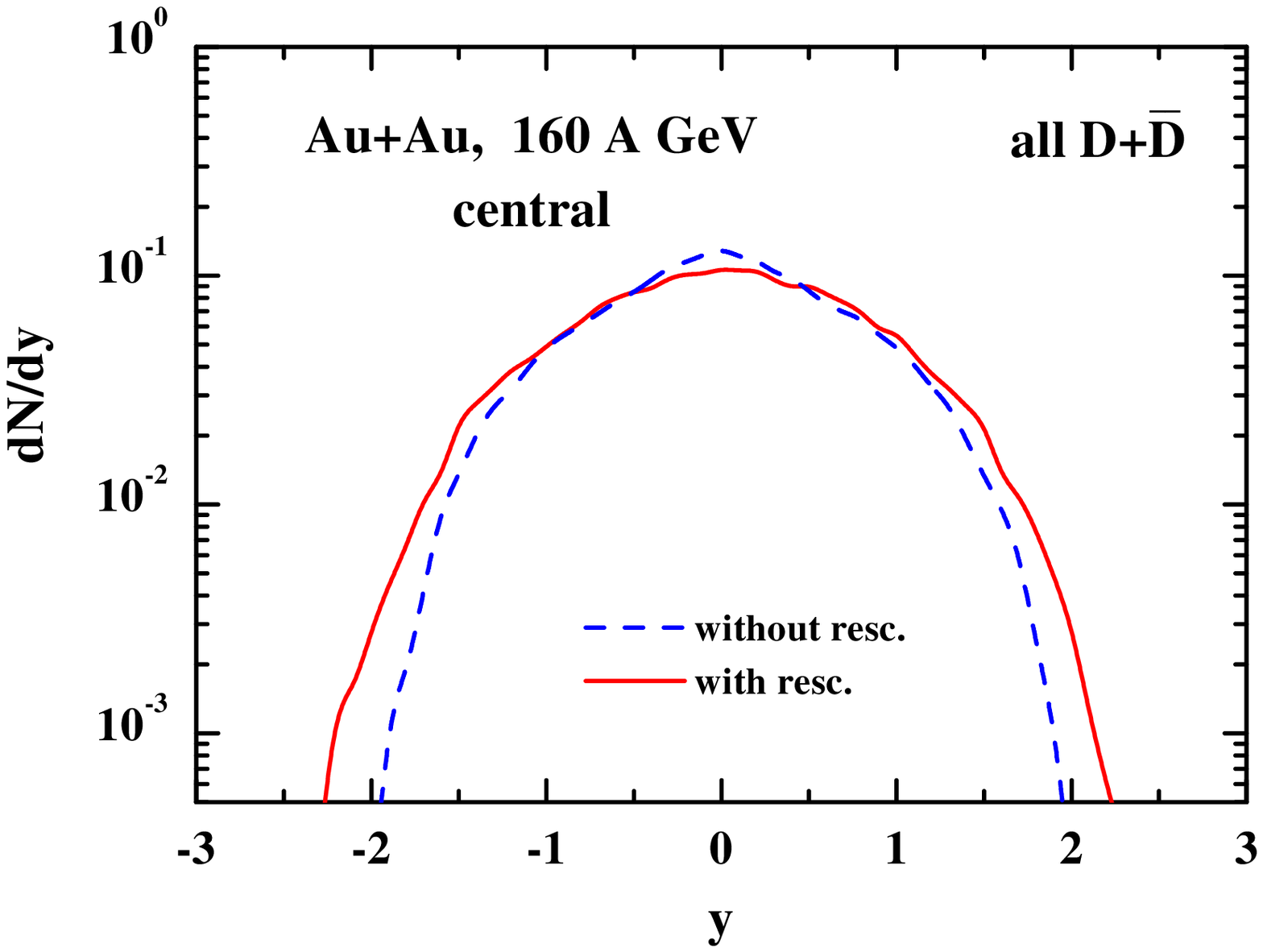,width=15cm}}
\vspace*{5mm}
\caption{The rapidity distribution of $D+\bar{D}$ mesons  in the HSD
approach for a central $Au+Au$ collision at 160 A$\cdot$GeV.  The dashed line
denotes the spectrum without rescattering and reabsorption while the
solid line includes the final state interactions of $D$-mesons with
hadrons.}
\label{bild11}
\end{figure}

\begin{figure}[h]
\centerline{\psfig{file=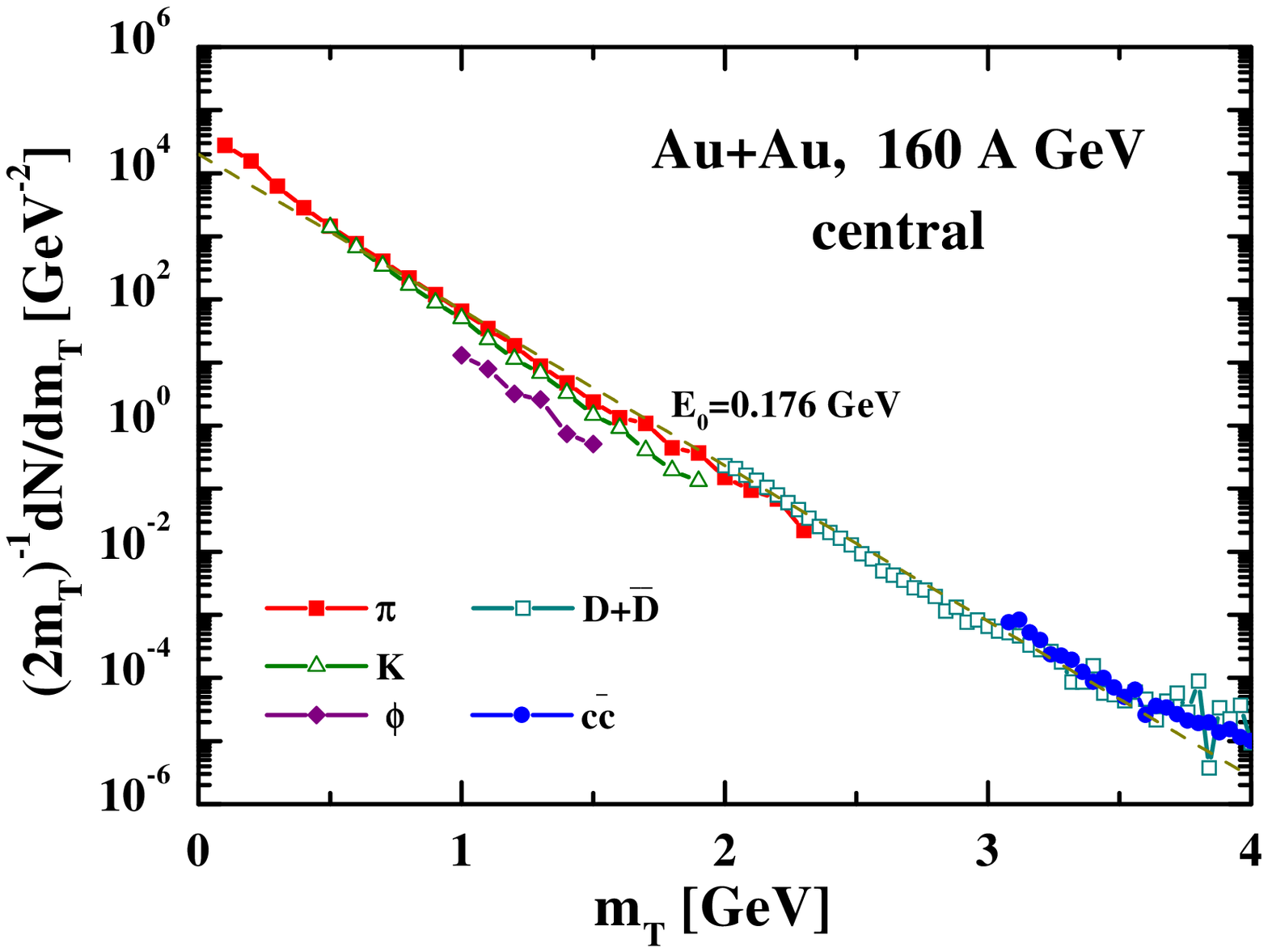,width=15cm}}
\vspace*{5mm}
\caption{The transverse mass spectra of pions (full squares), kaons
(open triangles), $\phi$-mesons (full rhombes), $D+\bar{D}$ mesons
(open squares) and $J/\Psi, \Psi^\prime$ mesons (full dots) in the HSD
approach for a central $Au+Au$ collision at 160 A$\cdot$GeV. The thin
dashed line shows an exponential with slope parameter $E_0$ = 0.176
GeV. Note that final state elastic scattering of kaons and
$\phi$-mesons with pions has been discarded in the calculations. }
\label{bild12} \end{figure}

\begin{figure}[h]
\centerline{\psfig{file=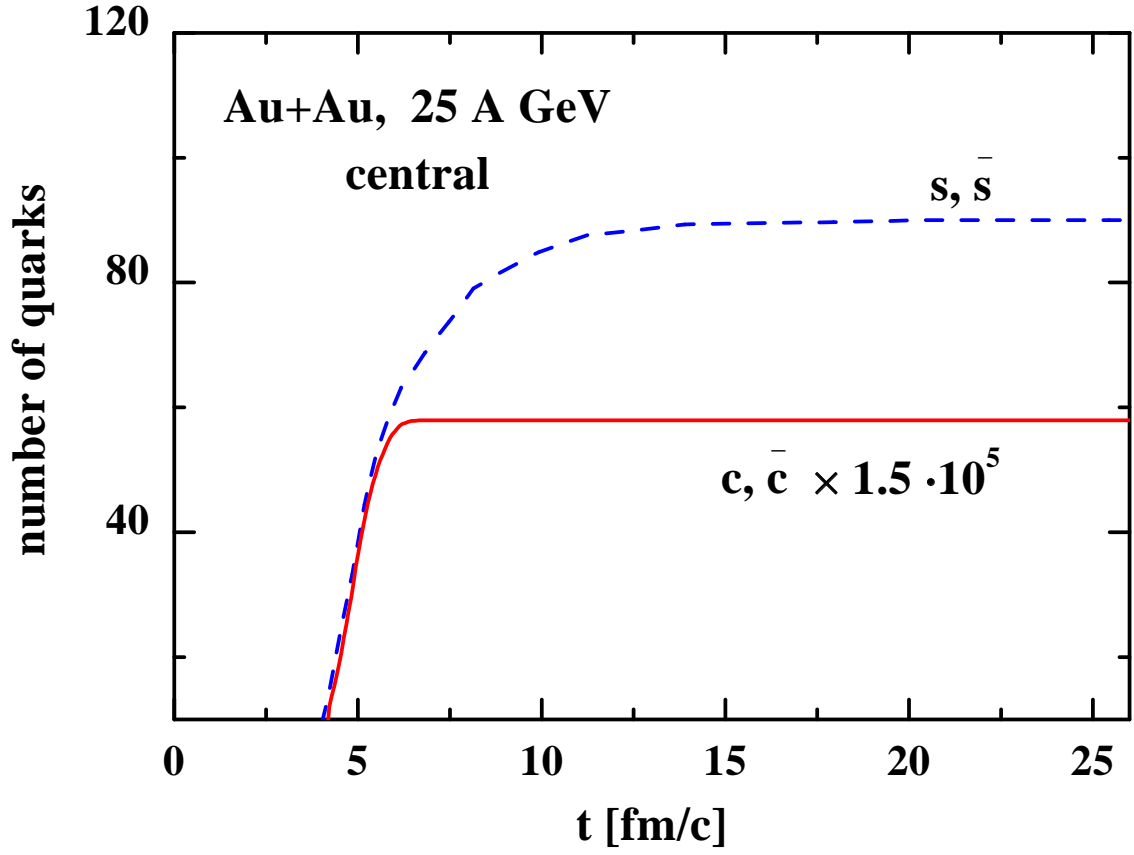,width=15cm}}
\vspace*{5mm}
\caption{Time evolution for the production of $s, \bar{s}$ (dashed
line) and $c \bar{c}$ quarks (solid line) in the HSD approach for a
central $Au+Au$ reaction at 25 A$\cdot$GeV.  The $c, \bar{c}$ numbers have been
scaled to the initial hard scattering processes by a factor of $1.5 \times 10^5$.}
\label{bild13}
\end{figure}

\begin{figure}[h]
\centerline{\psfig{file=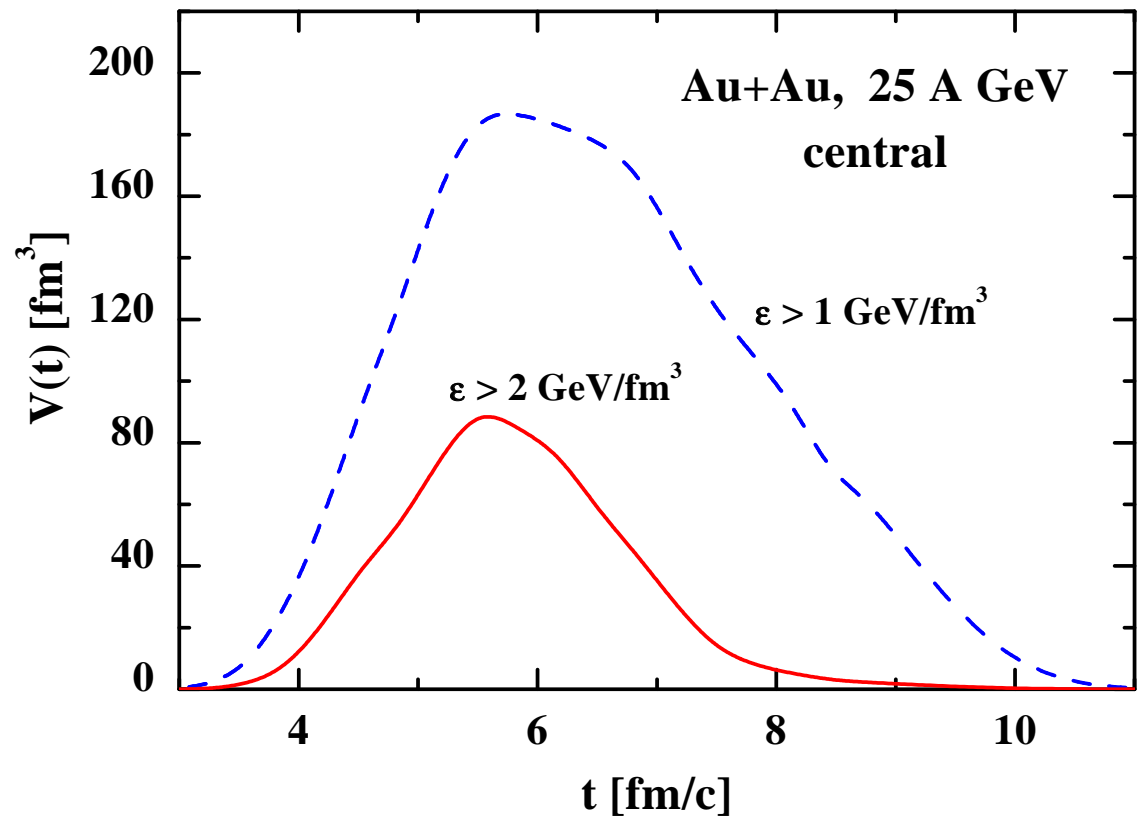,width=15cm}}
\vspace*{5mm}
\caption{Time evolution for the volume with energy density $\varepsilon
\geq$ 1 GeV/fm$^3$ (dashed line) and $\geq$ 2 GeV/fm$^3$ (solid line) in the
HSD approach for a central $Au+Au$ reaction at 25 A$\cdot$GeV. }
\label{bild14}
\end{figure}

\begin{figure}[h]
\centerline{\psfig{file=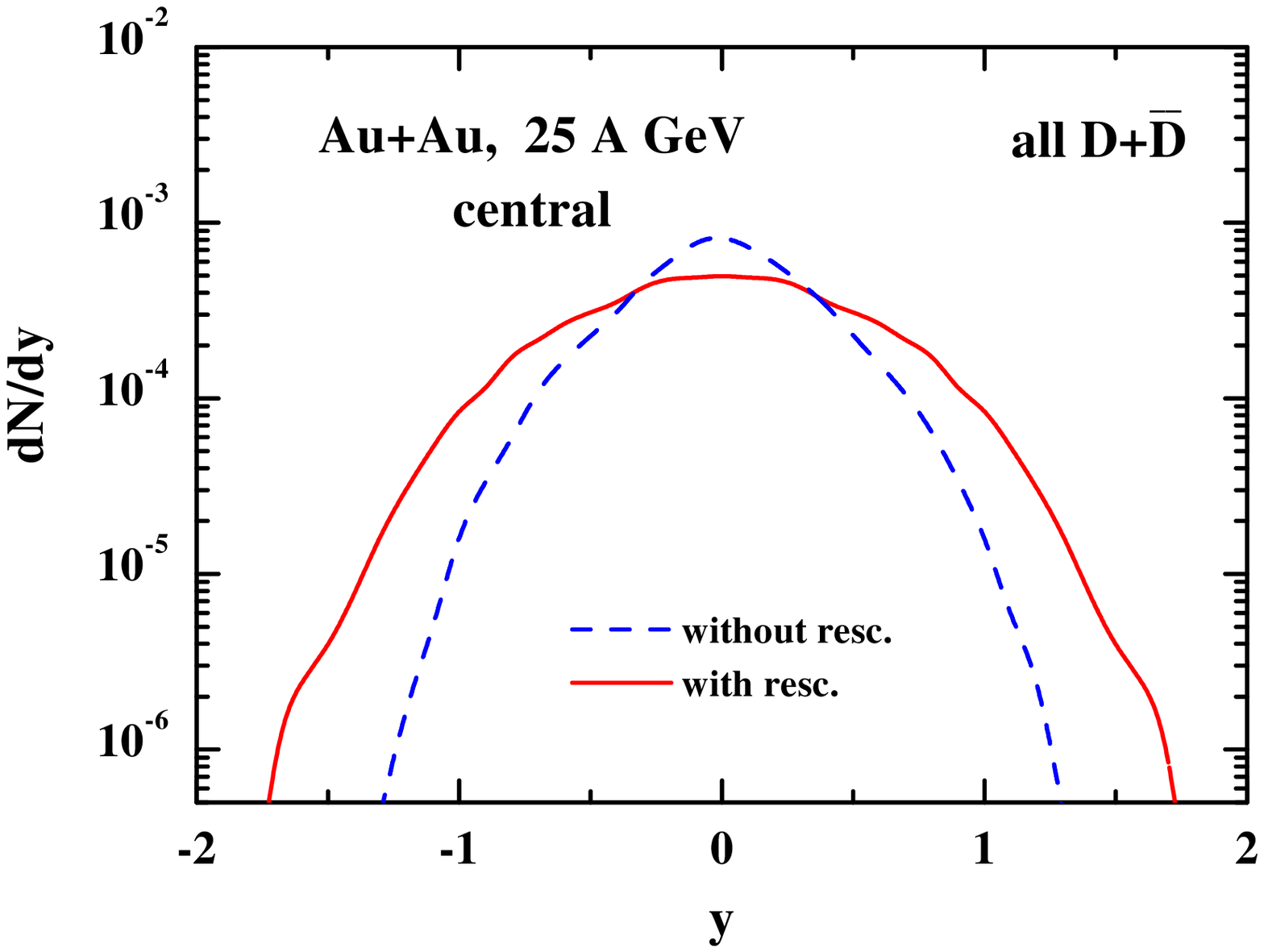,width=15cm}}
\vspace*{5mm}
\caption{The rapidity distribution of $D+\bar{D}$ mesons  in the HSD
approach for a central $Au+Au$ collision at 25 A$\cdot$GeV.  The dashed line
denotes the spectrum without rescattering  while the
solid line includes the final state interactions of $D$-mesons with
hadrons.}
\label{bild15}
\end{figure}

\begin{figure}[h]
\centerline{\psfig{file=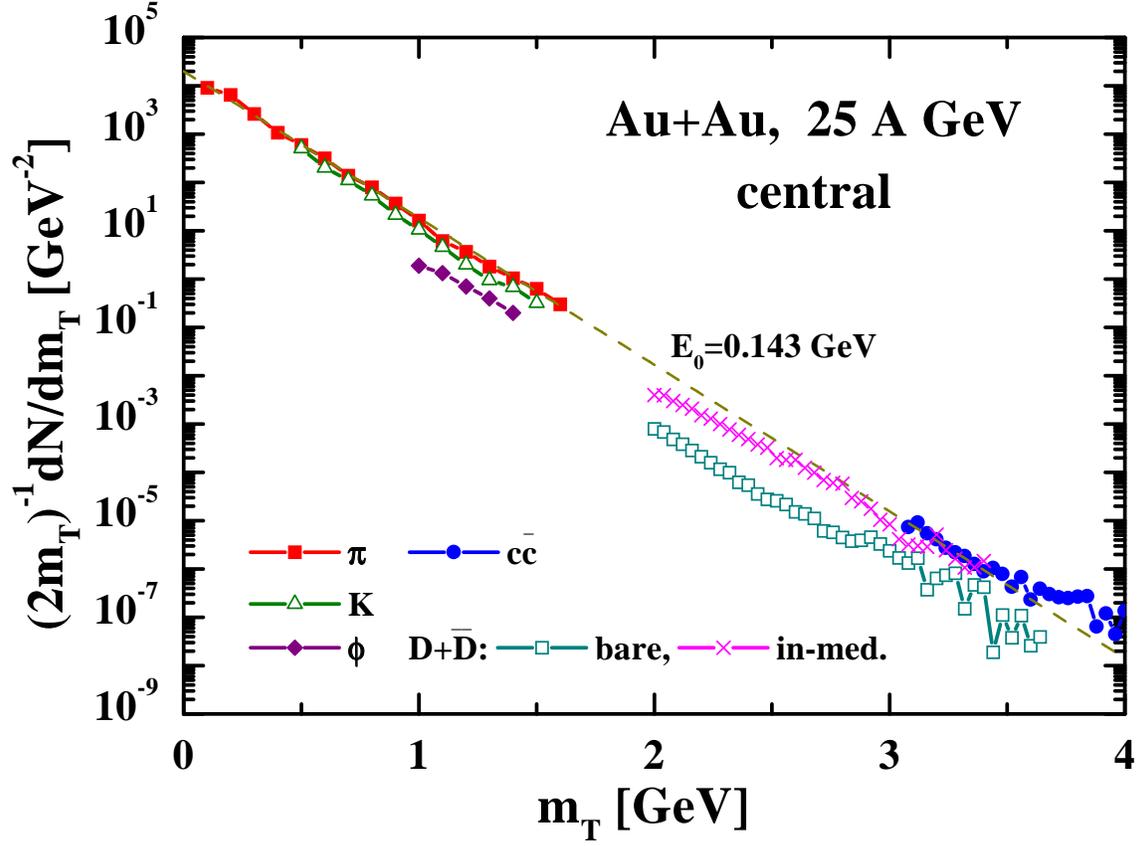,width=15cm}}
\vspace*{5mm}
\caption{The transverse mass spectra of pions (full squares), kaons
(open triangles), $\phi$-mesons (full rhombes),$D+\bar{D}$ mesons (open
squares) and $J/\Psi, \Psi^\prime$ mesons (full dots) in the HSD
approach for a central $Au+Au$ collision at 25 A$\cdot$GeV without
including self energies for the mesons. The crosses stand for the
$D$-meson $m_T$ spectra when including an attractive mass shift
according to (\protect\ref{drop}). The thin dashed line shows an
exponential with slope parameter $E_0$ = 0.143 GeV. Note that final
state elastic scattering of kaons and $\phi$-mesons with pions has been
discarded in the calculations.}
\label{bild16}
\end{figure}

\begin{figure}[h]
\centerline{\psfig{file=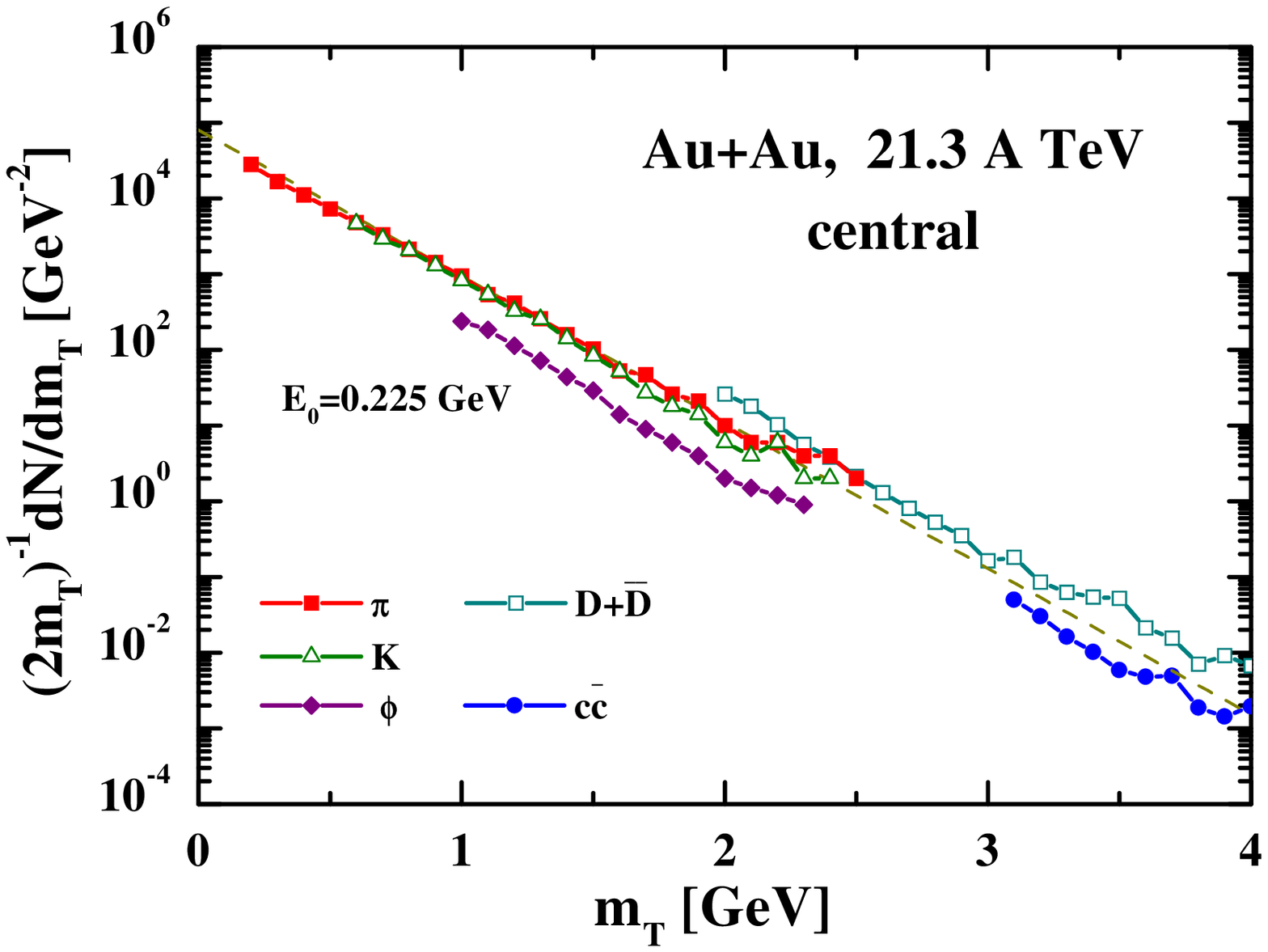,width=15cm}}
\vspace*{5mm}
\caption{The transverse mass spectra of pions (full squares), kaons
(open triangles), $\phi$-mesons (full rhombes),$D+\bar{D}$ mesons (open
squares) and $J/\Psi, \Psi^\prime$ mesons (full dots) in the HSD
approach for a central $Au+Au$ collision at 21.3 A$\cdot$TeV. The thin
dashed line shows an exponential with slope parameter $E_0$ = 0.225
GeV. Note that final state elastic scattering of kaons and
$\phi$-mesons with pions has been discarded in the calculations.}
\label{bild16b}
\end{figure}

\begin{figure}[h]
\centerline{\psfig{file=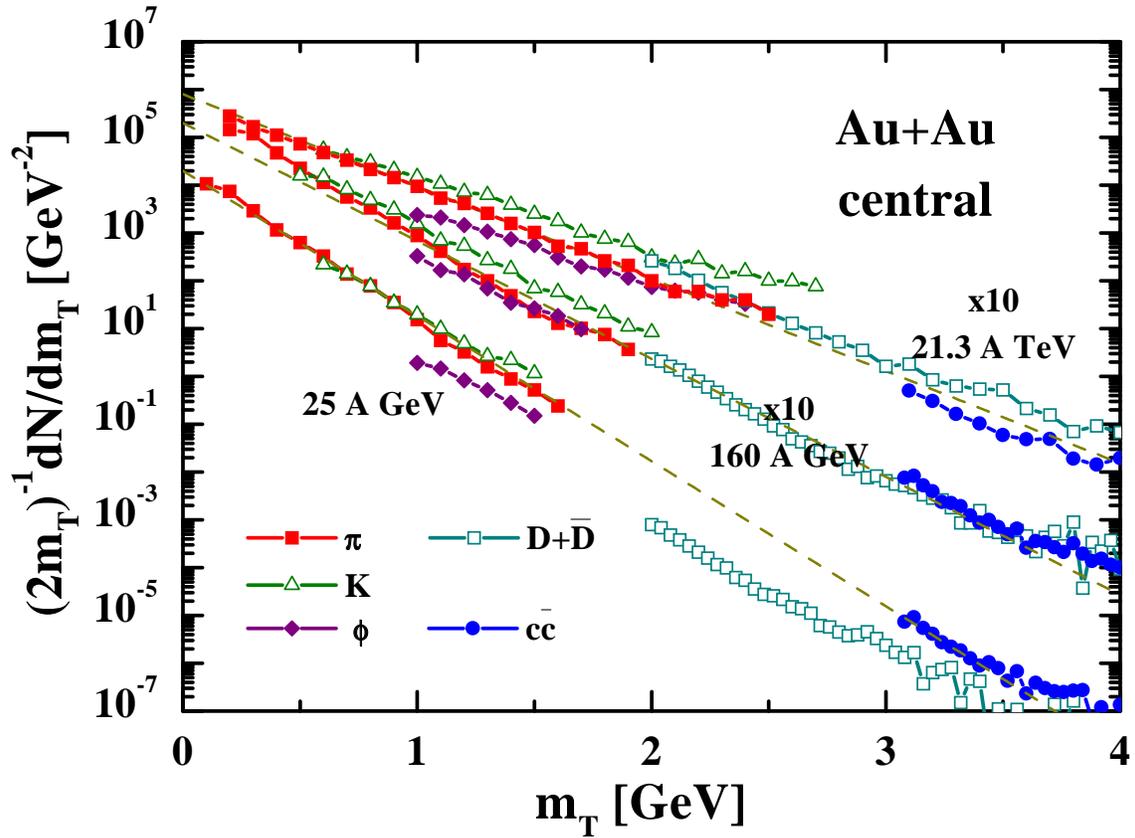,width=15cm}}
\vspace*{5mm}
\caption{The transverse mass spectra of pions (full squares), kaons
(open triangles), $\phi$-mesons (full rhombes),$D+\bar{D}$ mesons (open
squares) and $J/\Psi, \Psi^\prime$ mesons (full dots) in the HSD
approach for  central $Au+Au$ collisions at 25 A$\cdot$GeV, 160
A$\cdot$GeV and 21.3 A$\cdot$TeV. The spectra at 160 A$\cdot$GeV and
21.3 A$\cdot$TeV have been multiplied by a factor of 10.  The thin
dashed lines show exponentials with slope parameters $E_0$ = 0.143,
0.176 and 0.225 GeV, respectively. The deviations from the thin lines
essentially are due to final state elastic scattering of kaons and
$\phi$-mesons with pions, which accelerate the heavier mesons (see
text). }
\label{bild16c}
\end{figure}

\begin{figure}[h]
\centerline{\psfig{file=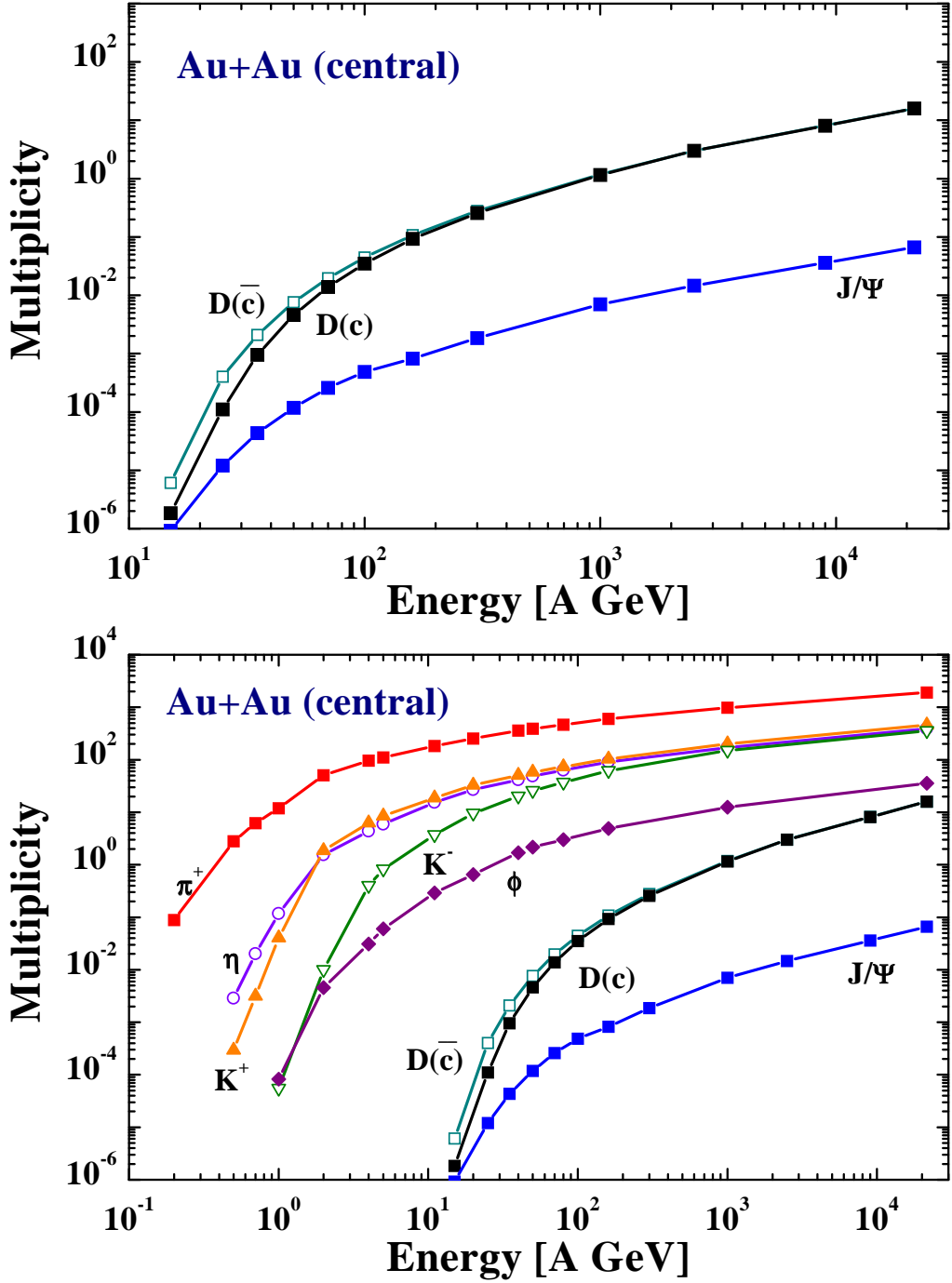,width=13.5cm}}
\vspace*{5mm}
\caption{The multiplicity of $D,\bar{D}$ and $J/\Psi$-mesons (upper
part) for central collisions of $Au+Au$ in the HSD approach including
elastic and inelastic reactions, but no in-medium modifications of
their spectral functions. The multiplicities for $\pi^+, \eta, K^+,
K^-$ and $\phi$ (in the lower part) have been taken from Ref.
\protect\cite{Cass00} while the lines for $D(c), D(\bar{c})$ and
$J/\Psi$ are the same as in the upper part.}
\label{bild17}
\end{figure}

\end{document}